# Network structure and patterns of information diversity on Twitter

Jesse Shore, Jiye Baek and Chrysanthos Dellarocas

Boston University Questrom School of Business

**ABSTRACT**

Social media have great potential to support diverse information sharing, but there is widespread concern that platforms like Twitter do not result in communication between those who hold contradictory viewpoints. Because users can choose whom to follow, prior research suggests that social media users exist in "echo chambers" or become polarized. We seek evidence of this in a complete cross section of hyperlinks posted on Twitter, using previously validated measures of the political slant of news sources to study information diversity. Contrary to prediction, we find that the average account posts links to more politically moderate news sources than the ones they receive in their own feed. However, members of a tiny network core do exhibit cross-sectional evidence of polarization and are responsible for the majority of tweets received overall due to their popularity and activity, which could explain the widespread perception of polarization on social media.

# Network structure and patterns of information diversity on Twitter

## 1. INTRODUCTION

Because anyone can post and re-share content, social media has been connected to increased participation and diversity of expression, raising hopes for a role for social media in promoting innovation, building social capital and empowering workers within firms and in society in general (An et al., 2011; Bertot, et al., 2010; Kane et. al, 2009; Woodly, 2008). Given the opportunities available in big data and the imperative to make use of them (LaValle, et al, 2013), business leaders have turned in increasing numbers to analyzing social media data in order to learn from customers (Culnan, McHugh, and Zubillaga, 2010; He, Zha and Li, 2013; Chen, Chiang, and Storey, 2012), and computer scientists have developed many tools to help achieve these ends (see e.g. Pang and Lee, 2008). Firms have also adopted internal social networking platforms in great numbers. Yammer, a popular enterprise social networking platform, claims to be used by more than 500,000 firms, including 85% of the Fortune 500 (Yammer, 2015).

It is easy to see why many see social media as potentially valuable external sources and internal conduits of diverse knowledge (Kane, Majchrzak and Ives, 2010). Innovation has long been seen as deriving from recombining diverse ideas (Schumpeter, 1934), and diverse ideas are assumed to flow through diverse networks (Hampton, Lee, and Her, 2011) like those created by connecting a diverse user base via social media. In general, diversity among individuals is thought to lead to better performance in solving problems (Hong and Page, 2004) and modern crowdsourcing approaches to innovation would seem to thrive on the fuel of diversity (Jeppesen and Lakhani, 2010).

However, information technologies, while providing historically unprecedented potential for free public expression, also provide self-regulating mechanisms that allow users to customize content feeds. In making these choices, people tend to connect with similar others (McPherson, Smith-Lovin, and Cook, 2001) and seek out information that confirms their previously-held beliefs (Nickerson, 1998). It is therefore unclear if the diverse points of view of social media users ever actually come into contact with each other, or if they cyber-balkanize themselves into "echo chambers" in which they are only exposed to ideas they already hold (Van Alstyne and Brynjolffson, 2005).

Many of the most lucid and powerful research studies on this topic to date have been in the setting of political information diversity and communication – a setting we also study in the present paper.  In addition to being economically and societally consequential, political communication is an appealing setting for the study of information diversity: there is a clear left-right spectrum of opinion, which simplifies the difficult issue of how to measure diversity in a meaningful way (Page, 2010).  Additionally, it is not too much of a stretch to view political communication among social media users in the United States as an example of a market for information in which two principal organizations (the Democratic and Republican political parties) are competing for attention and influence.

In prior literature, there is some evidence in favor of a tendency to echo chambers, some evidence in favor of polarization and still other evidence in favor of a tendency for people to limit themselves to expression of moderate and mainstream ideas on social media (see discussion in Section 2).  We believe that a likely reason for the conflicted nature of the literature is that earlier work has generally focused too narrowly on unrepresentative or incomplete data sets. Social networks and online communities often have a core-periphery structure consisting of a highly interconnected core of important and active nodes, surrounded by a larger, less densely connected periphery (Borgatti and Everett, 2000; Dahlander and Fredriksen, 2012, Wu, et al., 2011).

By focusing on highly active users, prior research on the phenomena of echo chambers and polarization has arguably only emphasized the study of the network core, whose behavior is not representative of the average user of the platform (Adamic and Glance, 2005; Conover, et al, 2011; Bakshy, Messing and Adamic, 2015). Moreover, it could even be argued that by constructing their data sets by including only those individuals with clear partisan affiliation (Adamic and Glance, 2005; Bakshy, Messing and Adamic, 2015), or those who posted about politically divisive topics (Conover, et al., 2011; Barbera et al., 2015), prior research studied only users prone to political division and therefore sheds little light on the nature of social media in general. Due to its traditional survey methodology, Hampton, et al. (2014) does not have this limitation but on the other hand it also cannot answer those questions which would require large-scale network data as evidence.

Here, we seek to reconcile the differing perspectives on patterns of diversity in social media with a study of a nearly complete cross-section of Twitter posts (“tweets”) containing hyperlinks to news stories, together with the associated follower network data. Our data set includes 908,000 tweets posted by 215,174 users based on a 300 hour data set, representing a nearly complete record of all such activity on Twitter during the collection period. We test hypotheses implied by prior research as well as characterize the overall structure of the Twitter follower network with respect to ideological diversity. Because we have the follower graph, we are also able to relate the slant of the information an account receives to the slant of the information they post themselves. Rather than echo chambers or cross-sectional evidence of polarization, we find that, on average, Twitter accounts post links to more politically moderate (but not necessarily centrist) news sources than the links they receive in their own feed. Members of a tiny but highly followed network core behave differently from the typical user, however, and post links to sources that are more politically extreme than what they receive in their own newsfeeds. While our empirical setting is political slant, we believe that the implications go beyond this narrow application and provide a basis for understanding the structure of self-organization in social media more generally.

## 2. THEORIES OF INFORMATION DIVERSITY ON SOCIAL MEDIA

No one can read every article or interact with every user on the internet; instead, internet users must make choices about where to direct their attention. Given the human tendency toward homophily (McPherson, Smith-Lovin and Cook, 2001) and confirmation bias (Nickerson, 1998), social media users are likely to follow other users whose opinions are similar to their own. At the extreme, this could lead to fragmentation of users into ideologically narrow groups, in which people are only exposed to information that confirms their previously-held opinions (Van Alstyne and Brynjolffson, 2005; Burt, 2004). We refer to this as the "echo chambers" theory of social media. Empirical studies have confirmed some of these fears: there is a tendency for blogs with the same political and ideological inclination to link to each other (Adamic and Glance, 2005; Conover, et al, 2011; Hargittai, Gallo and Kane, 2008) and a tendency of readers to engage with a small subset of content (Schmidt, et al., 2017) aligned with their ideological preferences (Lawrence, Sides and Farrell, 2010). Homophilous behavior is then magnified by algorithmic information filters on certain social media sites such as Facebook (Bakshy, Messing and Adamic, 2015; Lazer, 2015).

A related view says that homophily may not lead people to be disconnected and ignorant of opposing views, as echo chambers theory would have it. Instead, the relationship between groups of connected individuals may be mutually aware and antagonistic. Sunstein (2002, 2008) argues that when like-minded individuals discuss a controversial topic, there is a tendency for them to adopt an even more extreme position on that topic than they initially held. Barbera et al. (2015) document this process unfolding over time in partisan debate of controversial issues on Twitter. Conover et al. (2011) show that while people "retweet"[1] like-minded others on Twitter, they "@-mention"[2] users they disagree with in the context of argument and other negative commentary. The prevalence of @-mentions illustrates that although conservatives and liberals do not retweet each other's content, they are indeed aware of each other and may even follow each other. In short, on Twitter, people belong to groups that are antagonistic, not ignorant of each other. We refer to this as the "polarization" theory of social media.

Despite all of this evidence, however, the idea that social media users segregate themselves into homogenous or polarized communities is far from an established fact. Some have theorized that while social network ties may tend to be formed among similar others, there are many dimensions along which that similarity may be manifest (Watts, Dodds, and Newman, 2002) and social media users may be connected not only to people with whom they agree politically, but also to people with whom they share other similarities, such as workplace, alma mater and so on. This phenomenon of simultaneous contact with people from different contexts has been called "context collapse" and can lead users to limit their expression of potentially controversial beliefs (Marwick and boyd, 2010; see also Bernstein (2012) for a similar finding in an organizational context). Centola and Macy (2007) argue that certain phenomena – including potentially controversial expressions such as political beliefs – are most likely to occur and exert influence in the context of a highly clustered network such that there is the possibility of receiving multiple reinforcing signals from one's network neighbors. These recent theories echo the pre-internet theory of public opinion that people tend to articulate what they perceive to be the mainstream point of view or withhold their voice entirely, creating a "spiral of silence" for minority viewpoints (Noelle-Neumann, 1974). Finally, to the extent that individuals fear they are being surveilled by government actors, they may share more politically moderate views in public than those they actually hold (see Stoycheff, 2016). Collectively, we refer to these ideas as the "mainstreaming" theory of social media.

[1] To "retweet" is to re-share a message one has received with one's own followers

[2] On Twitter an "@-mention" is to include another user's handle in a tweet (post), prepended with an @ sign, which uniquely identifies the specific individual, creates a clickable hyperlink to their profile page and notifies the target individual that they have been mentioned in the tweet.

There is empirical support for the mainstreaming narrative of social media use as well. On average, it has been found that people are much less likely to discuss controversial topics on social media than in private (Hampton et al. 2014). For political hashtags on Twitter, repeated exposures are important precursors to an individual's adoption of those hashtags in their own posts (Romero, Meeder and Kleinberg, 2011), which could be interpreted as seeking repeated confirmation from their community before sharing something potentially controversial.

Our primary goal is to consider evidence for and against the theories of echo chambers, polarization, and mainstreaming. In this and the following sections, we therefore ask what we would expect to find in a complete cross-section of Twitter posts if the above theories were in fact true. We articulate a number of detailed hypotheses to test on this basis, but our overarching questions are simply whether Twitter shows evidence of widespread (1) echo chambers (2) polarization and/or (3) mainstreaming. Before going into these specific theories, we state the relatively uncontroversial hypothesis that social media users display homophily. In other words, we would expect the typical Twitter user to tweet links to news sources with similar political slant to the slant of the content they receive from the people they follow: we expect followers and followees to tweet at a similar level of political slant (we define how we measure slant below).

> *Hypothesis 1H: The mean political slant of news sources in tweets by individuals is significantly correlated to the mean political slant of the tweets that they receive from their followees.*

For notational convenience, we group differing hypotheses about the same variables under the same hypothesis number, but append letters to indicate which theories they correspond to. For instance, *Hypothesis 1H* describes the homophily hypothesis about the relationship between incoming and outgoing slant, while *Hypothesis 1EC, Hypothesis 1P,* and *Hypothesis 1M* (below) describe the echo chambers hypothesis, polarization hypothesis and mainstreaming hypothesis, respectively. We continue to use "H" for homophily, "EC" for echo chambers, "P" for polarization, and "M" for mainstreaming throughout.

### 2.1 Echo Chambers

For our purposes, what cross-sectional observations would be consistent with echo chambers? In an echo chamber, one only hears things one might have said oneself. If the homophily of Hypothesis 1H is strong enough to create echo chambers, we would expect not just correlation between political slants, but indeed for people to tweet at the same mean level of political slant as those that they follow.

> *Hypothesis 1EC: The mean political slant of news sources in tweets sent by individuals is the same as the mean political slant of the tweets that they receive from the people they follow.*

In addition to mean outgoing slant being the same as mean incoming political slant, a meaningful echo chamber would lack diversity in information, in which people send and receive information from only a narrow range of viewpoints. This lack of diversity drives conformity with a slanted perspective. By the same logic, an individual who receives diverse information would tweet in a more politically balanced way:

> *Hypothesis 2EC: The larger the standard deviation of political slant an account receives, the closer mean outgoing slant will be to the political center*

The imagery of "echo chambers" further suggests a closed social network neighborhood, in which ideas reverberate among people who know each other but who are disconnected from the outside world. Network theory supports the notion that people in highly clustered positions in the social network are likely to be similar with respect to political slant: individuals in such dense clusters accrue shared, mutual knowledge as a consequence of communicating with each other (Granovetter, 1973; Hansen, 1999; Burt, 2004). Moreover, beliefs and social behaviors spread by the process of 'complex contagion,' (Centola, 2010) in which the 'infectiousness' of a behavior increases with repeated exposures and thus tend to spread within rather than between clusters of ties (see also Romero, Meeder and Kleinberg, 2011 for evidence on political speech on Twitter consistent with the theory of complex contagions). As a result, we would expect people within dense clusters to be more politically similar to each other than people who are not in highly clustered network positions.

> *Hypothesis 3EC: The greater the clustering around an individual, the closer the political slant in their own tweets is to the political slant in the tweets they receive from the people they follow.*

## 2.2 Polarization

Sunstein (2002, 2008) argues that when like-minded individuals are connected, the views they express can be more extreme than what they would have expressed prior to deliberation, in part because of social pressure toward conformity. He refers to this phenomenon as polarization. Conover et al. (2011) document polarization as a phenomenon in which the two sides are not ignorant of each other (as the simple echo chambers theory would have it) but rather in conflict: although people do not retweet information from users on the other side of the political spectrum, they are aware of them and what they are saying. In this view of polarization (that we adopt here), people read at least some information from both sides of the political spectrum, but only tweet out information consistent with their own side. If Twitter accounts are not just homophilous but also polarized, it would suggest that they tweet at more extreme levels of slant than the average slant of information they receive in their news feeds (which, because it contains at least some information from the other side of the spectrum is closer to the political center than the slant of the information consumed from their own side of the spectrum).

*Hypothesis 1P: The mean political slant of news sources in tweets by individuals is more extreme than the mean political slant of the tweets that they receive from the people they follow.*

If exposure to opposing viewpoints brings polarization, then indeed this theory makes a different prediction from that of Hypothesis 2EC with respect to the diversity of incoming ideas:

*Hypothesis 2P: The larger the standard deviation of political slant an account receives, the more extreme outgoing slant is relative to incoming slant.*

## 2.3 Mainstreaming Theory

Finally, we turn to hypotheses implied by mainstreaming. The component theories that make up our "mainstreaming" theory (especially context collapse and the spiral of silence) do not explicitly make predictions about the choices individuals make about what information to consume – only the information they choose to put into the public domain themselves. A very strong hypothesis that takes the original idea of a spiral of silence very literally is that whatever they read, we would find people tweeting only politically centrist content themselves.

However, to successfully tweet at the true political center, people would have to know where the political center was. This may not be realistic. Instead, people's understanding of which opinions are centrist may be biased toward the information they are exposed to (e.g. democrats will think the center is further left than republicans will). Indeed, it has been shown that centrist content is often perceived as biased against one's own side (Vallone, Ross and Lepper, 1985). Therefore, a more behaviorally realistic version of the mainstreaming hypothesis is that we would expect individuals to Tweet material that is more politically neutral (centrist) than what they receive, even if it is not at the true political center.

> *Hypothesis 1M: The mean political slant of news sources linked to in an individual's own tweets is more politically centrist than the mean political slant in the tweets they receive from the people they follow.*

## 2.4 Beyond average behavior: macroscopic and subnetwork analyses

The above hypotheses are specified in microscopic terms, —in that we treat individuals as the units of analysis—and will be tested on the entirety of link-posting behavior on Twitter during the study period. These hypothesis tests serve our theoretical questions and provide the foundation of our empirical analysis.

To paint a fuller picture, however, and to better connect our work with prior research on social media, we also include a series of analyses that take other perspectives on the data. In particular, scholars of online communities have been concerned with their macroscopic core-periphery structure (Dahlander and Fredriksen, 2012; Collier and Kraut, 2012; Wasko, Teigland and Faraj, 2009), which Wu, et al. (2011) have demonstrated also describes Twitter networks. In a classic core-periphery structure, the network core is a set of nodes (individuals) that tend to be connected to each other; the periphery is a (typically larger) set of nodes that tend to be connected to nodes in the core, but not to each other (Borgatti and Everett, 1999). In the setting of Twitter, this is to say that there is a set of highly-followed accounts (the core) that tend to follow each other; more typical users (the periphery) follow members of the core, but are less likely to follow other typical users.

### 2.4.1 Analyses of macroscopic structure

Members of the news-sharing core differ from other users with respect to their network position; it is possible that they also differ from other users in terms of the correlation between incoming slant and outgoing slant. To check for this possibility, we repeat a basic analysis of the relationship between incoming and outgoing slant on two subgraphs[3] of the Twitter news-sharing network. In particular, we distinguish those accounts that are highly followed and active in posting many links to news items from those that are not.

We expect to find a higher correlation between incoming and outgoing slant in the 'news-centric core'—the subgraph of individuals who are both highly followed and post many news items—than in subgraphs comprising individuals who are not highly followed or do not post many news items or both. People in the news-centric core may be maintaining a public identity centered around news, and so may connect with fewer people for reasons other than discussion of news. They may also engage in self-conscious management (Marwick and boyd, 2010) of their list of followees -- to demonstrate party loyalty, for example -- which would result in a higher correlation between incoming and outgoing slant.

[3] A subgraph consists of a certain subset of nodes of a larger network, along with all of the links between them.

In contrast to members of the news-centric core, those who are highly followed but do not post many links to news items are probably highly followed for other reasons, such as celebrity, and may not pay as much attention to the variable of political slant when choosing whom to follow. Those who post many links to news items but are not highly followed may well demonstrate homophily on political slant, but because they are less likely to be public figures, they may not be curating their followee list as self-consciously as members of the news-centric core. Finally those who are neither highly followed (among the individuals in our data comprising people who posted hyperlinks and their followers and followees) nor highly active posters of news may be less active users of Twitter, or actively using Twitter for other purposes, and thus are expected to demonstrate less homophily on political slant than those in the news-centric core.

*Hypothesis 4H: The mean political slant of news sources in tweets sent by individuals in the news-centric core of Twitter users is closer to the mean political slant of the tweets that they receive from their followees than for people outside of the news-centric core.*

Following the logic above and Wu, et al.'s (2011) observation that 'coreness' is not a binary but rather a continuous variable, we would expect that the higher the thresholds we use to separate individuals who are "highly followed" and "post many links to news" from everybody else (i.e. the stricter the definition of what constitutions the news-centric core), the more homophilous members of the core would be. Moreover, prior work (Conover, et al., 2011) has found not just homophily but indeed polarization when focusing on accounts in what we argue is likely to be the network core.

*Hypothesis 4P: The stricter the definition of what constitutes the news-centric core, the more extreme outgoing slant is, relative to incoming slant.*

### 2.4.2 Correspondence of macroscopic network structure and political slant

Earlier studies of political division on social media have shown a clear correspondence between the macroscopic structure of a network and the political slant of its nodes (Adamic and Glance, 2005; Conover et al, 2011). In particular, this work shows that the network is starkly divided into two (one liberal and one conservative) modular clusters of nodes, such that nodes tend to be connected within each cluster, but only sparsely connected between clusters. By showing how cleanly political slant corresponds to network structure, these excellent studies lend strong support to the cyber-balkanization theory (Adamic and Glance, 2005) and polarization theory (Conover et al., 2011), discussed above.

These studies are nevertheless limited in two important ways. First, both studies use only a binary, liberal v. conservative representation of slant, preventing more nuanced examination of homophily. Second, both studies only consider the behavior of elites and self-identified partisans (i.e. members of the news-centric core, whom we have just argued are not representative of the typical user) and thus shed no light on how social media works as a platform for discourse for the vast majority of users.

Because our data includes a continuous representation of political slant and includes all Twitter users rather than only elites, we are able to address these two limitations. First, we are able to analyze the core separately from other users. Second, rather than considering *classifications* of nodes into two categories (liberal v. conservative), we focus instead on *permutations* of nodes defined either by political slant or other means.

A permutation is simply an ordering of the nodes of a network such that each node is assigned an ordinal number from 1 to N (where N is the number of nodes in the network). Permutations can be defined by any number of means, but in the present context we will be particularly interested in permutations derived from the political slant variables: those in which the nodes are ordered from most liberal (and thus given the number 1) to the most conservative (and thus given the number N). A "good" permutation is one in which nodes that are close together in the network are close together in the ordering of nodes.

The echo chambers and polarization theories have been associated with a close correspondence between network structure and political slant, while the mainstreaming theory predicts the opposite. In the following hypotheses (labeled 5Ma – 5Md, with an "M" to indicate that they express the mainstreaming point of view, and "a", "b", "c" or "d" to identify a specific detailed hypotheses), we compare the quality of permutations in this sense. Section 3.4.3, below, provides more concrete details on measurement of permutation quality.

Essentially, just as prior work (Adamic and Glance, 2005; Conover et al, 2011) showed that classifying nodes into liberal v. conservative was a good fit to the macroscopic division of the network into two distinct communities, we will ask if a continuous measure of political slant is a good one-dimensional description of network structure. We wish to ask this question for both incoming and outgoing slant and for the network core and network periphery. To establish whether permutations based on incoming or outgoing slant are "good" descriptions of macroscopic network structure, we will compare them to permutations derived from standard community discovery algorithms (see methods, below); we repeat this process separately for the core and the periphery.

Accordingly, we test the following hypotheses about the quality of nodal permutations for the core.

*Hypothesis 5Ma: Outgoing political slant is not a good description of the community structure of the network core.*

*Hypothesis 5Mb: Incoming political slant is not a good description of the community structure of the network core.*

We go on to test the following hypotheses about the quality of nodal permutations for the periphery.

*Hypothesis 5Mc: Outgoing political slant is not a good description of the community structure of the network periphery.*

*Hypothesis 5Md: Incoming political slant is not a good description of the community structure of the network periphery.*

## 3. DATA AND METHODS

Our primary analysis is based on relatively old (from 2009), but exceptionally complete data that is described here. Additionally, to provide robustness checks, we conducted similar analyses on sampled data from early 2017. Data, methods and results of these robustness checks are presented in Appendix A.

### 3.1 The Twitter Dataset

Our Twitter data for our primary analysis comes from Galuba et al. (2010). For 300 continuous hours, starting on Thursday, September 10th, 2009, 19:56:47 GMT, the Twitter Search API was continuously queried for the search string "http". The text of each tweet returned by the query was parsed for any URLs and user names it contained. Each URL mentioned in the tweets was stored. If the URL was created by one of the popular URL shortening services (e.g. bit.ly), HTTP redirects were recursively followed to expand the URL to its original form. All the URLs were also URL-decoded to ensure uniform representation under the percent-encoding (%xx) notation. For each tweet, the Twitter API was queried for the metadata about the tweet's author as well as all the users that the author follows.

### 3.2 Measurement of Political Slant

Gentzkow and Shapiro (2011) published measurements of the political slant of the 119 most widely visited sources of online news in the United States, building on data from comScore Plan Metrix with 12 months data in 2009 (the same year as the Twitter data). Plan Metrix data come from a survey distributed electronically to approximately 12,000 comScore panelists. The survey asks panelists the question "In terms of your political outlook, do you think of yourself as. . .? [very conservative / somewhat conservative/ middle-of-the-road/ somewhat liberal / very liberal]". The average number of daily unique visitors in each category is reported by comScore for each site for each month.

Using this data, they posit the model of utility of a visit to a website in equation 1. The utility is that of user *i* going to site *j* on visit *k* on a given day, given the site quality $\alpha$, political slant $\gamma$, and dummy variable *c* set to 1 if visitor *i* is conservative and 0 if they are liberal (they omit data from individuals who answered "middle of the road").

$$u_{ijk} = \alpha_j + (2c_i - 1)\gamma_j + \varepsilon_{ijk} \qquad (1)$$

They estimate these parameters from the visit data, under the discrete choice modeling assumption that the visit would be made if and only if $u_{ijk} \geq u_{irk}\ \forall r \neq j$. We use the estimated parameter $\gamma$ as our measure of political slant. We also use $\alpha$ as a control variable indicating site quality. For the analysis, we use all tweets that contain any of the 119 domain URLs from Gentzkow and Shapiro (2011)[4].

Although Plan Metrix data are only available for relatively large sites, visits to news sites are highly concentrated. The 119 sites in the sample represent over 95% of all visits to news sites via independent browsing online (Gentzkow and Shapiro, 2011), and given the greater expected concentration of exposure on social media than independent browsing (i.e. finding news stories through other means, such as direct access of sites, search engines, or news aggregators) (Hong, 2012), the sample is expected to be even more completely representative for the setting of Twitter.

Because the data set for our primary analysis is relatively old, we also conducted a robustness check on 2017 data. This required us to calculate updated measurements of political slant and devise an approach to constructing an informative, if much smaller, sample of Twitter accounts. These methods, along with results for robustness checks are reported in Appendix A.

### 3.3 Variables

For each user, we calculated mean and standard deviation of incoming political slant (incoming slant) and mean and standard deviation of outgoing source slant (outgoing slant). The distribution of an individual account's outgoing slant comprises the political slant of each URL source that the user tweeted. The distribution of incoming slant comprises the slant score of every URL tweeted by the individuals whom that user follows (his/her followees). For both incoming and outgoing slant, if a news source was tweeted more than once, its slant score would

[4] for further detail on Genztkow and Shapiro's methods, see their online appendix: http://web.stanford.edu/~gentzkow/research/echo_modelappdx.pdf

be counted more than once in the distribution. Similarly, we calculate mean incoming (outgoing) quality for each individual from Gentzkow and Shapiro's $\alpha$. Finally we tabulate the count (number) of incoming and outgoing tweets for each user. Note that we fit models on data for users that both sent and received tweets containing links to news sources; to calculate incoming slant, however, we consider tweets from all users, including those who did not receive any news links in their own timelines. The output of those Twitter users who are widely followed but do not follow other accounts (typically public figures) is therefore still accounted for in the data.

Using the Twitter data, we construct a follower-followee network. A directed network tie exists from user *i* to user *j* if user *j* is a follower of user *i*. From this data, we calculated the clustering coefficient, (Watts and Strogatz, 1998), which captures the degree to which one's followers and followees also follow each other. More specifically, it is a measure of how many links there are among a node's neighbors, divided by the number of links that could exist among a node's neighbors[5].

## 3.4 Statistical Models

### 3.4.1 Average behavior

We fit ordinary least squares (OLS) models to test all versions of Hypotheses 1-3. For the variants of Hypothesis 1, the primary independent variable is incoming slant. Hypotheses 2 and 3 concern how one variable affects the relationship between incoming and outgoing slant. Thus, for variants of Hypothesis 2, we are interested in the interaction between the standard

[5] Let **A** denote the adjacency matrix of a network, such that $\mathbf{A}_{ij} = 1$ when there is a tie from node *i* to node *j*. Further, let $nei(i) = \{j: \mathbf{A}_{ij} = 1 \vee \mathbf{A}_{ji} = 1\}$ denote the set of vertices that are neighbors of node *i*, and $g_i = |nei(i)|$ denote the number of neighbors of node *i*. Then the clustering coefficient of node i, $C_i$ is defined as follows.

$$C_i = \frac{\sum_j \sum_k \boldsymbol{A}_{jk} = 1,\ j,k\ \in nei(i)}{(g_i)(g_i - 1)}$$

deviation of incoming slant and the main effect[6] of incoming slant. For Hypothesis 3EC, we are interested in the interaction between clustering coefficient and the main effect of incoming slant.

### 3.4.2 Core-Periphery structure

Hypotheses 4H and 4P concern the difference between the behavior of people who are highly followed and post many news articles and other individuals. To test these, we select nodes that are greater than or equal to some threshold quantiles, *s* and *t*, of outdegree and number of news stories posted, for $s, t \in \{0.75, 0.80, 0.85, 0.95\}$. We then consider the induced subgraph containing only those nodes that have outdegree greater than *s%* of nodes and have posted more news stories than *t%* of nodes in the full network. We then regress outgoing slant (dependent variable) on incoming slant (explanatory variable) for only those tweets coming from within this subgraph and separately, for all tweets coming from any source using OLS.

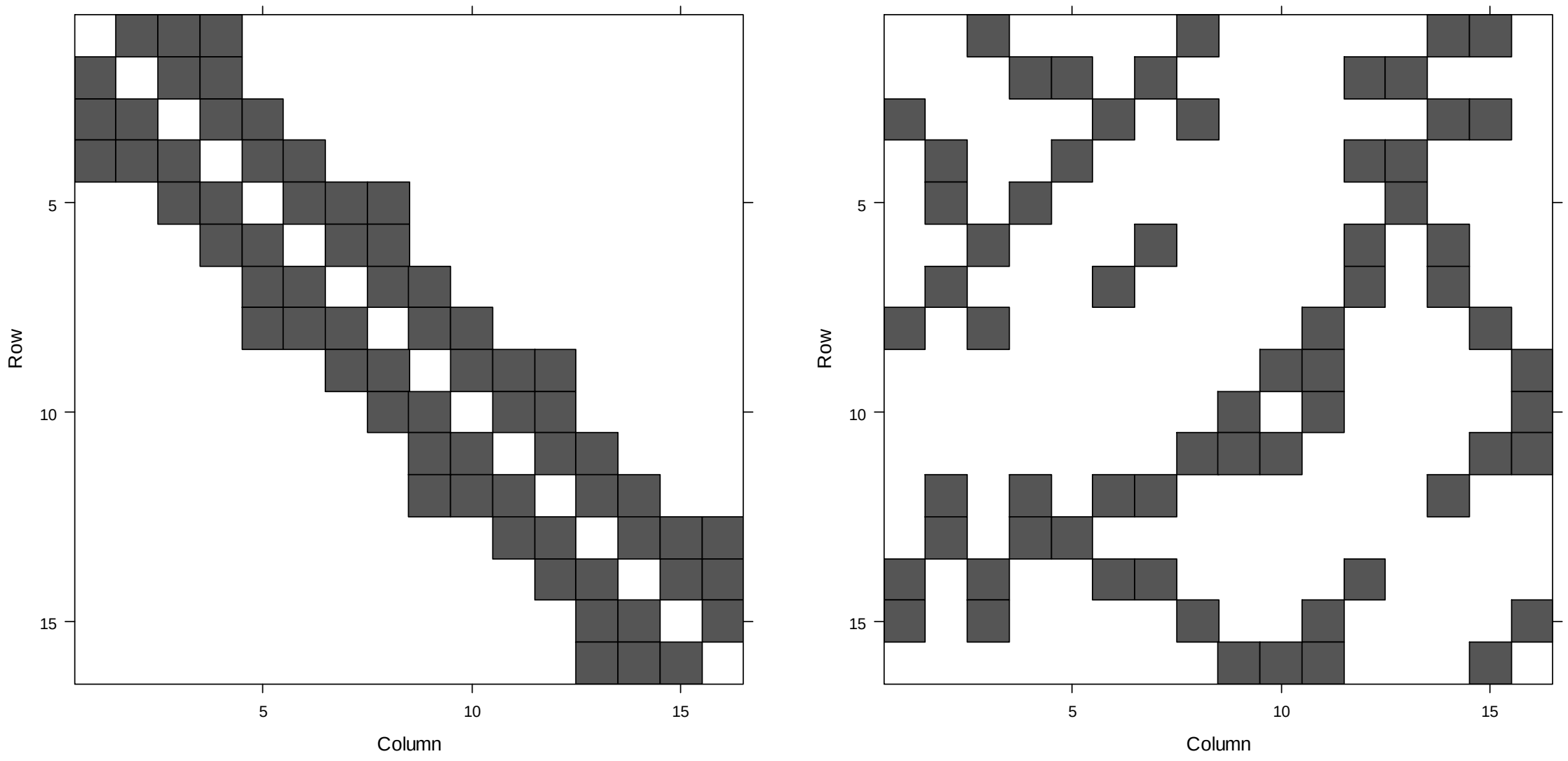

**Figure 1**: An illustration of two different permutations of the same matrix. Left: a good permutation that places connected nodes close together in the permutation and as a result concentrates ties toward the diagonal of the matrix. Right: a bad permutation, in which nodes that are connected in the network are not close in the permutation; as a result, tie weight is not concentrated toward the diagonal.

[6] NB: here and below "effect" is intended only in the sense of "statistical effect" and not in the sense of "causal effect."

### 3.4.3 Concordance of community structure and slant

Hypotheses 5Ma-5Md concern the concordance between the macroscopic community structure of the network and the political slant of the nodes. For these hypotheses, we are asking if permutations based on slant are good in the sense that nodes that are closely connected in the network are also close together in the permutation ordering (see Figure 1 for a toy example).

However, it is unclear a priori how to measure such correspondence between slant and structure, and then, how to determine if a given level of correspondence between slant and structure is a lot of correspondence or only a little. In other words, how good is good? In order to test these hypotheses, we therefore (1) define a measure of permutation quality (2) use standard community-discovery algorithms from the literature to define permutations that represent network structure well and measure their quality, (3) measure the quality of permutations based on slant, and (4) define a significance test to determine if the quality of the slant-based permutations are significantly worse than the community-discovery algorithmic permutations.

Hypotheses 5Ma, and 5Mb concern the news-centric core, and Hypotheses 5Mc and 5Md concern those outside of the news-centric core. The "core" subgraph is defined as above, using nodes greater than or equal to some quantiles *s*, *t* of outdegree and news posting activity such that a regression of outgoing slant on incoming slant yields the highest estimated parameter. Because of the computational expense of conducting these analyses on all ~213,000 nodes outside of the news-centric core using our methods, we test the latter hypotheses on a subgraph consisting only of moderate users. We define this subgraph as giant component of those accounts between the 25$^{th}$ and 75$^{th}$ percentiles for outdegree and less than the 75$^{th}$ percentile for number of news items posted.  This results in a subgraph of 75,640 Twitter accounts (a little more than one third of all news-active accounts), which omits the large number of least active and least followed accounts.

To measure quality of permutations, we start with the intuitions that connected nodes (those that follow each other) should be close together in a "good" permutation and that ties (matrix entries equal to 1) between nodes whose indices are close together in a given permutation will be close to the diagonal of the permuted adjacency matrix. Conversely, of course, ties between nodes that are far apart in the permutation will be far from the diagonal in the permuted adjacency matrix.  We use these intuitions to define an idealized model against which to compare the observed permuted data such that we can evaluate them quantitatively.

We define the idealized model as a probability matrix, Z, such that matrix entries (network ties) closest to the diagonal are modeled as having probability 1, with linearly declining probability further from the diagonal.  Note that this is not a fitted model, so the "probabilities" are not estimated from the data; as an idealized model, the matrix of probabilities functions more as a "scoring matrix:" we calculate the likelihood of the idealized model, Z, under the observed permuted data in question.

Concretely,

$Z_{i,j} = 0$ (No self loops)

$Z_{i,j} = 1\ ,\ \ j \in \{i+1, i-1\}$ (Highest probability closest to diagonal)

$Z_{i,j} = Z_{i,j-1} - 1/n\ ,\ \ j \geq i+2$

(Decreasing probability with distance from diagonal)

$Z_{i,j} = Z_{i,j+1} - 1/n\ ,\ \ j \leq i-2$

where $i$ and $j$ are row and column indices, respectively, and $n$ is the number of nodes. In general, a log likelihood is calculated by summing the log probabilities of the data points, conditional on the model.  A matrix of probabilities of ties existing where they do in the permuted matrix P, conditional on Z can be obtained by simply taking the Hadamard (pointwise) product of Z and P. To calculate the log likelihood $\ell_p$ of Z under some permutated adjacency matrix, $P$, we then simply sum the log of the values in the resulting product matrix:

$$\ell_p = \sum_{ij} \ln(Z_{ij} P_{ij}) \quad (2)$$

The higher the log likelihood, the more closely the permuted observed matrix adheres to the idealized model, Z.

In addition to the permutations implicit in sorting the nodes according to their outgoing and incoming slants, we also consider two algorithmically-defined permutations deriving from the network structure (pattern of follower/followee ties) alone, rather than taking into account political slant or any other nodal attribute. Algorithmically defined permutations attempt to place nodes close together in the permutation ordering if they are close together in the network. In the first of these, we follow the usual procedure of spectral clustering: we calculate the eigenvectors of the Laplacian matrix (a transformation of the adjacency matrix representation of the network) and then rank nodes according to the values in the eigenvector corresponding to one of the smallest eigenvalues not equal to zero[7] (see e.g. (Dhillon, 2001; Von Luxburg, 2007) for more detail). To find the smallest eigenvalues and corresponding eigenvectors of the "moderate user" subgraph, we use ARPACK numerical methods (Lehoucq, Sorensen, and Yang, 1998), which are therefore approximate. In the second algorithmically-defined permutation, we use the method of Clauset, Newman and Moore (2004), as implemented in the igraph analytical software package (Csardi and Nepusz, 2006), which produces a full hierarchical dendrogram as a side effect of finding a smaller number of communities. We simply take the ordering of nodes at the bottom level of that dendrogram as our permutation.[8]

[7] Usually the best description of the macroscopic structure of a network is found in values of the eigenvector corresponding to the smallest non-zero eigenvalue, but not always. We therefore consider the eigenvectors corresponding to the 5 smallest eigenvalues and take the best, where best is defined as yielding the highest likelihood of $Z$ (see below).

We visualize the difference between these four permutations of nodes (two by slant and two by community discovery algorithm) by plotting the adjacency matrix, with the rows and columns in permutation order for the core and moderate users subgraphs (Figures 2 and 3). On these plots, if rows are indexed by $i \in \{1 \ldots N\}$, and columns are indexed by $j \in \{1 \ldots N\}$, then a point at location (*i*,*j*) on the visualization indicates that there exists a tie between node *i* and node *j* (account *j* follows account *i* on Twitter). The closer a permutation is to the idealized model, *Z*, the more the points in these plots will be concentrated toward the matrix diagonal.

It remains to be determined how much higher a log likelihood has to be to be considered significantly better than the likelihood of an alternative permutation of the observed matrix. Typically, likelihoods are compared via likelihood ratio tests. Strictly speaking, however, likelihoods calculated from a matrix probability model on two different permutations of the same data are not from nested models, and thus the chi-squared limiting distribution on the traditional likelihood ratio test cannot be assumed. Instead, we calculate a critical value for distinguishing between the likelihoods of this model under these two permutations computationally.

We calculate the worst-case reduction of log likelihood due to incorrect ordering for each of 5% of the total number of nodes and tabulate the reduction of log likelihood that would occur if the edges incident to those nodes were moved as far away from the diagonal as possible. We repeat this procedure 1000 times and take the 95% percentile of the resulting distribution to be the critical value. If the log likelihood of an alternative permutation is less than this critical value, we would consider that alternative permutation to be a worse description of network structure.

[8] We grant that this permutation is based on a partial, rather than full ordering of nodes, since the first pair of nodes that are grouped together in a given branch of the dendrogram could appear in either order in the final permutation. However, since the number of such interchangeable pairs is small, and the distance that each node in these pairs could move in the permutation is at maximum 1 spot, we take the partial ordering output from the R function to be representative of the quality of all such possible permutations.

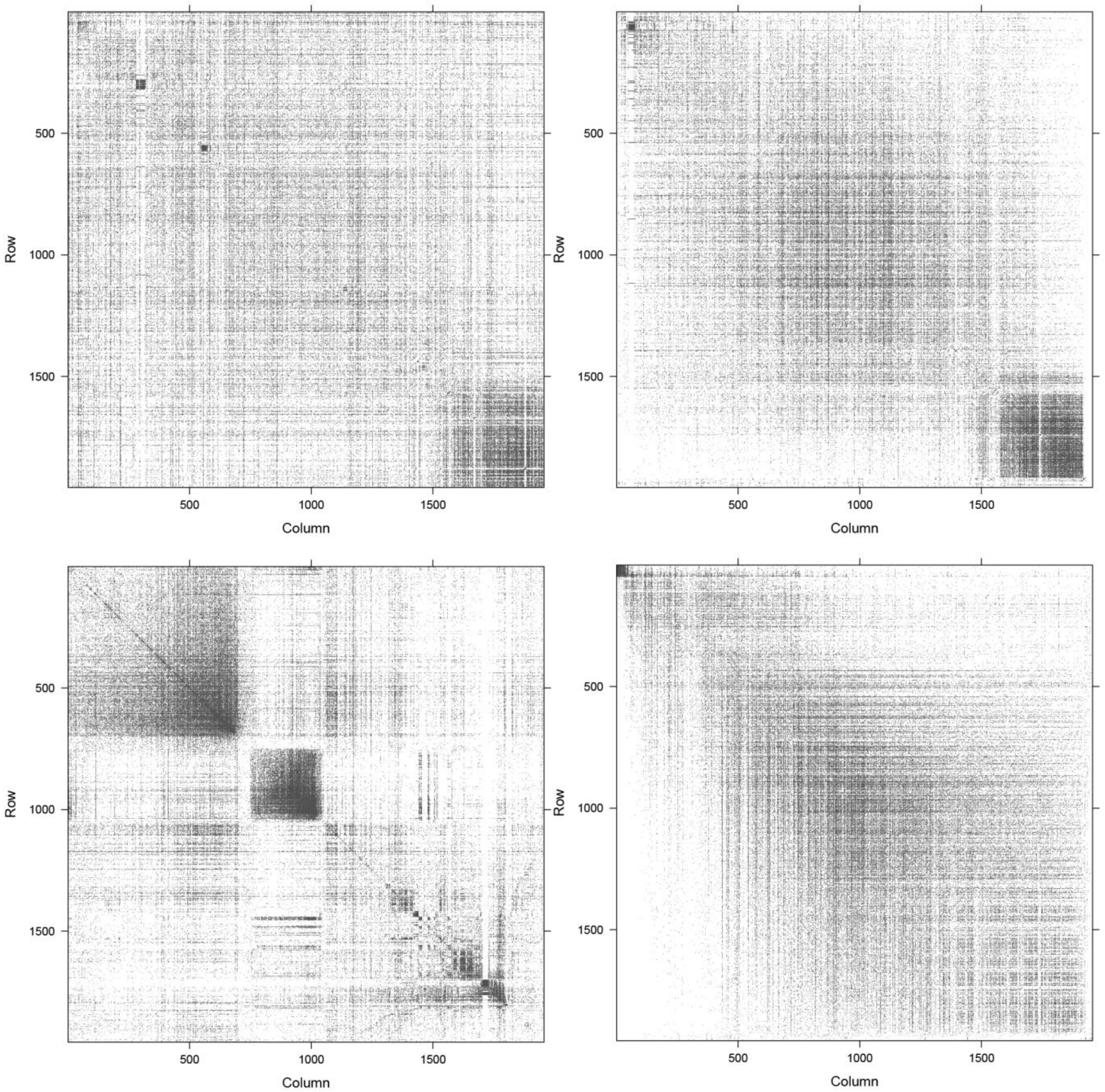

**Figure 2**: The adjacency matrix of the news-centric core, permuted by outgoing slant (top left) by incoming slant (top right) by the method of Clauset, Newman and Moore (2004) (bottom left) and by the values of the eigenvector corresponding to one of the smallest eigenvalues of the Laplacian matrix (bottom right).

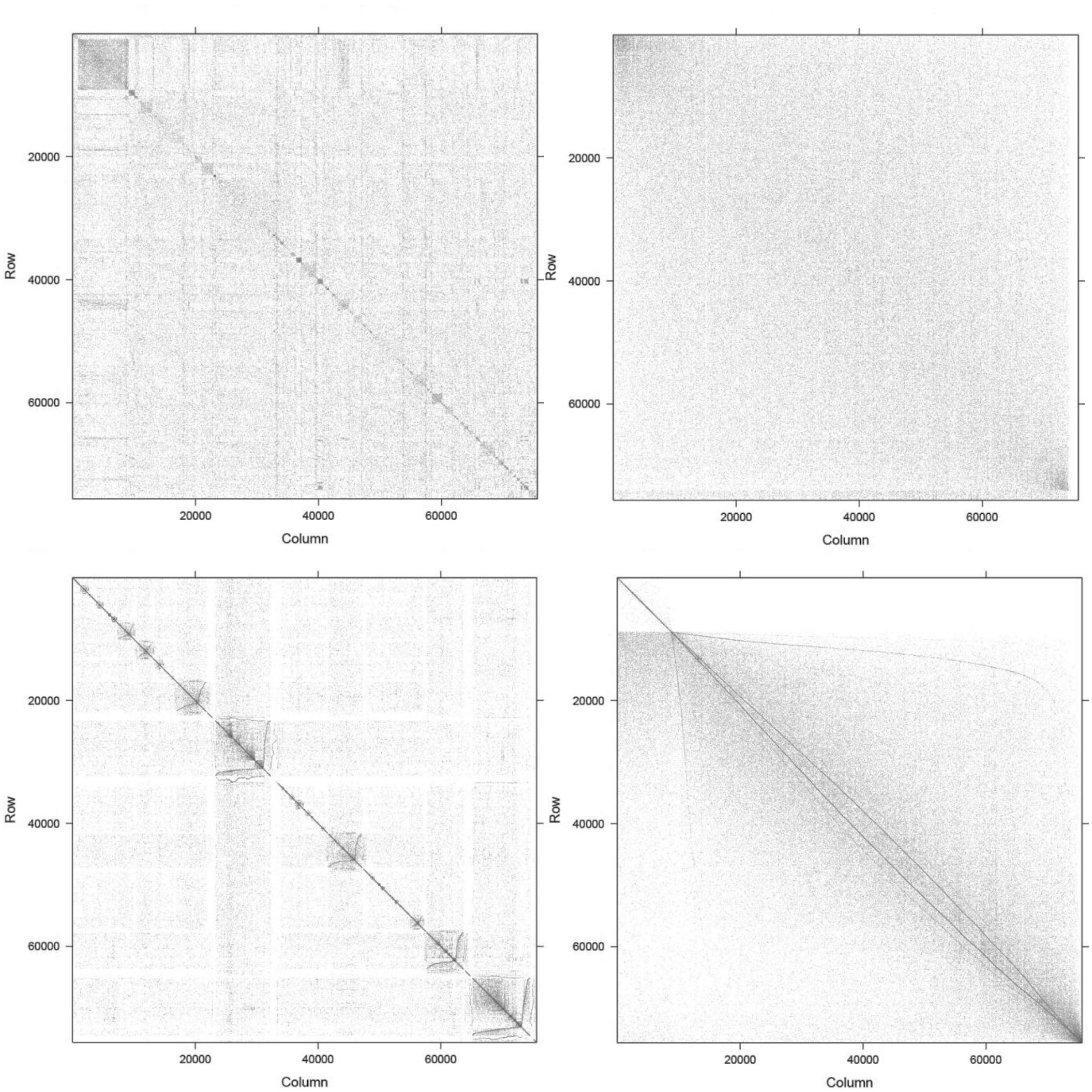

**Figure 3**: The adjacency matrix of the “moderate users” subgraph taken from the periphery, permuted by outgoing slant (top left) by incoming slant (top right) by the method of Clauset, Newman and Moore (2004) (bottom left) and by the values of the eigenvector corresponding to one of the smallest eigenvalues of the Laplacian matrix (bottom right).

### 3.5 Descriptive Statistics

In our dataset, after processing and selecting those who both received and sent tweets containing links to the sources we covered, we were left with a group of 215,174 Twitter accounts that posted 908,565 tweets containing a hyperlink to one of the 119 news sources for which Gentzkow and Shapiro provide an estimated political slant. There were 14,870,199 follower-followee relationships among these accounts, and only 6709 accounts did not follow and were not followed by any of the other accounts that posted links to the 119 domains.

There were 165,624 accounts that had outgoing slant less than zero (liberal) and 49,550 accounts that had outgoing slant greater than zero (conservative). This is consistent with Pew's survey results, which indicate that liberals were significantly more active on social media at that time (Pew Research Center, 2012). Descriptive statistics are in Table 1 and correlations are in Table 2. We also tabulated the counts of users by mean incoming slant and mean outgoing slant. As Table 3 shows, we find that some people read tweets from the opposite side of the political spectrum from the side they tweet on themselves.

Table 1: Descriptive Statistics

| | Min | Median | Mean | Max | Sd |
|---|---|---|---|---|---|
| Mean outgoing slant | -1.5568 | -0.251 | -0.233 | 2.263 | 0.393 |
| Count of outgoing tweets | 1 | 1 | 4.156 | 3321 | 22.495 |
| Quality of outgoing tweets | 0 | 5.970 | 5.802 | 8.630 | 1.251 |
| Mean incoming slant | -1.557 | -0.233 | -.226 | 1.879 | 0.232 |
| Count of incoming tweets | 0 | 147 | 715.8 | 100984 | 2238.6 |
| Quality of incoming tweets | 0 | 5.878 | 5.847 | 8.630 | 0.768 |
| Standard Dev. of incoming slant | 0 | 0.335 | 0.321 | 1.687 | 0.139 |
| Count of outgoing retweets | 0 | 0 | 0.7631 | 867 | 3.96 |
| Outdegree | 0 | 11 | 70.77 | 52731 | 348.7 |
| Indegree | 0 | 17 | 71.33 | 32511 | 252.66 |
| Clustering Coefficient | 0 | 0.081 | 0.108 | 1 | 0.116 |

Table 2: Correlation matrix

| | Mean outgoing Slant | Count of outgoing tweets | Quality of outgoing tweets | Mean incoming Slant | Count of incoming tweets | Quality of incoming tweets | Outdegree | indegree | Clustering Coefficient | Count of retweets | ln(outdegree+2)/ ln(indegree+2) |
|---|---|---|---|---|---|---|---|---|---|---|---|
| outSlant | 1 | 0.017 | -0.1342 | 0.395 | 0.063 | -0.087 | 0.025 | 0.0352 | 0.0163 | 0.0315 | 0.0276 |
| outCount | 0.017 | 1 | -0.005 | 0.034 | 0.086 | -0.015 | 0.0554 | 0.0463 | -0.0419 | 0.2044 | 0.1243 |
| outQuality | -0.134 | -0.005 | 1 | -0.084 | -0.023 | 0.243 | -0.0095 | -0.0154 | -0.0084 | -0.0221 | -0.0167 |
| inSlant | 0.395 | 0.034 | -0.0836 | 1 | 0.09 | -0.216 | 0.0413 | 0.0586 | 0.0351 | 0.0423 | 0.0278 |
| inCount | 0.063 | 0.086 | -0.023 | 0.09 | 1 | -0.033 | 0.6449 | 0.8956 | -0.1116 | 0.0893 | -0.0269 |
| inQual | -0.087 | -0.015 | 0.243 | -0.216 | -0.033 | 1 | -0.021 | -0.0345 | 0.033 | -0.0256 | -0.0818 |
| outdegree | 0.025 | 0.055 | -0.0095 | 0.041 | 0.645 | -0.021 | 1 | 0.7685 | -0.0992 | 0.0413 | 0.0943 |
| indegree | 0.035 | 0.046 | -0.0154 | 0.059 | 0.896 | -0.035 | 0.7685 | 1 | -0.1125 | 0.0537 | -0.0074 |
| Clustering Coef | 0.016 | -0.042 | -0.0084 | 0.035 | -0.112 | 0.033 | -0.0992 | -0.1125 | 1 | -0.0159 | -0.1747 |
| rtCount | 0.031 | 0.204 | -0.0221 | 0.042 | 0.089 | -0.026 | 0.0413 | 0.0537 | -0.0159 | 1 | -0.0011 |
| ln(OD+2)/ ln(ID+2) | 0.028 | 0.124 | -0.0167 | 0.028 | -0.027 | -0.082 | 0.0943 | -0.0074 | -0.1747 | -0.0011 | 1 |

Table 3: Tabulation of mean incoming and outgoing political slant by account

| | Mean incoming slant | | | | | | |
|---|---|---|---|---|---|---|---|
| Mean out. slant | (-1.75,-1.25] | (-1.25,-0.75] | (-0.75,-0.25] | (-0.25,0.25] | (0.25,0.75] | (0.75,1.25] | (1.25,1.75] |
| (1.75,2.25] | 0 | 0 | 9 | 38 | 18 | 2 | 0 |
| (1.25,1.75] | 0 | 2 | 29 | 100 | 71 | 6 | 0 |
| (0.75,1.25] | 0 | 25 | 1073 | 2847 | 1153 | 100 | 3 |
| (0.25,0.75] | 1 | 112 | 4959 | 11640 | 3335 | 123 | 4 |
| (-0.25,0.25] | 6 | 536 | 27949 | 47400 | 2432 | 125 | 7 |
| (-0.75,-0.25] | 4 | 519 | 43881 | 34305 | 973 | 67 | 1 |
| (-1.25,-0.75] | 2 | 1840 | 14856 | 7241 | 281 | 19 | 0 |
| (-1.75,-1.25] | 0 | 3 | 222 | 122 | 1 | 0 | 0 |

# 4. RESULTS

## 4.1 Average behavior

We begin with results for average behavior of all individuals who tweeted a link to one of the sites covered by the Genztkow and Shapiro data. We report regression coefficients from OLS models in Table 4. To ease interpretability with respect to the main variables of interest (incoming and outgoing slant), all other variables are centered at their mean and scaled to unit variance. Additionally, all covariates are included both as main effects and as interaction effects with incoming slant.

Most notably, the estimated parameter for the mean political slant of sites linked-to in incoming tweets ranged from 0.667 to 0.724, significantly less than 1. Robustness checks with recent data (see Appendix A) are in agreement with the full analysis presented here; parameter estimates from 2017 data were 0.586 and 0.751. Additional statistical results are mentioned in line with the text, below.

Hypothesis 1H stipulates that there is a significant correlation between the political slant in incoming versus outgoing tweets. As just mentioned above, we estimated a positive and significant regression parameter for this relationship across all models. We do find homophily, and the null hypothesis is therefore rejected.

**4.1.1 Echo Chambers**

Hypothesis 1EC makes a stronger statement than Hypothesis 1H about the relationship between incoming and outgoing slant, positing that they are equal. If Hypothesis 1EC were true, we would expect the regression parameter to be equal to 1.0 (meaning the outgoing slant is equal to 1.0 times the incoming slant, and therefore equal). Given the standard errors of the estimated coefficients, we fail to reject a composite null hypothesis of correlation but not equality: the outgoing slant is not equal to the incoming slant. The slant at which people tweet is correlated with but not equal to the slant of the material they receive.

Table 4 Relationship between incoming and outgoing political slant

| | DV: Mean slant of sites in outgoing tweets | | | |
|---|---|---|---|---|
| | I | II | III | IV |
| Mean slant, sites in incoming tweets | 0.667***<br>0.003 | 0.697***<br>0.004 | 0.720***<br>0.004 | 0.724***<br>0.004 |
| *ln*(Count of incoming tweets) | | 0.036***<br>0.001 | 0.032***<br>0.002 | 0.029***<br>0.004 |
| *ln*(Count of outgoing tweets) | | 0.010***<br>0.010 | 0.007***<br>0.001 | 0.007***<br>0.001 |
| Mean quality of sites in incoming tweets | | 0.008***<br>0.001 | 0.008***<br>0.001 | 0.008***<br>0.001 |
| Std deviation of slant of incoming tweets | | 0.013***<br>0.001 | 0.012***<br>0.001 | 0.012***<br>0.001 |
| *ln*(# followers+2) | | | 0.013***<br>0.002 | 0.018***<br>0..002 |
| *ln*(# followers+2) ÷ *ln*(#followees+2)† | | | 0.007***<br>0.001 | 0.007***<br>0.002 |
| Clustering coefficient | | | | 0.012***<br>0.001 |
| Incoming slant x ln(Count incoming tweets) | | 0.086***<br>0.004 | 0.018***<br>0.005 | -0.004<br>0.005 |
| Incoming slant x ln(Count outgoing tweets) | | 0.044***<br>0.003 | 0.040***<br>0.003 | 0.042***<br>0.003 |
| Incoming slant x quality of incoming tweets | | 0.023***<br>0.003 | 0.020***<br>0.003 | 0.022***<br>0.003 |
| Incoming slant x st. dev. of incoming slant | | 0.086***<br>0.003 | 0.066***<br>0.003 | 0.062***<br>0.003 |
| Incoming slant x *ln*(# followers+2) | | | 0.133***<br>0.007 | 0.156***<br>0.008 |
| Inc. slant x *ln*(#followers+2) ÷ *ln*(#followees+2)† | | | -0.058***<br>0.004 | -0.068***<br>0.004 |
| Incoming slant x Clustering coefficient | | | | 0.030***<br>0.003 |
| Intercept | -0.083***<br>0.001 | -0.084***<br>0.001 | -0.078***<br>0.001 | -0.078***<br>0.001 |
| # of Twitter accounts | 208465 | 208465 | 208465 | 204464 |
| Adjusted $R^2$ | 0.156 | 0.171 | 0.173 | 0.177 |

Notes: standard errors are printed below parameter estimates.
***: p< 0.001; **: p< 0.01; *: p < 0.05
Except for mean incoming slant and mean outgoing slant, all variables are centered and scaled to unit variance
† the logarithm of the number of followers/ees plus 2 is taken to avoid dividing by zero for those with no followees.

Hypothesis 2EC concerns the relationship between the diversity of news consumed and the effect of incoming slant on outgoing slant. The theory of echo chambers implies that less diversity should result in more conformity; in other words, given the fact that the estimated parameter for the main effect of incoming slant is less than one, we should expect a negative sign on the estimated parameter for the interaction between the standard deviation of incoming slant with the mean incoming slant, such that less diversity results in an outgoing slant closer to incoming slant. Contrary to the prediction of hypothesis 2EC, we do not see this; in fact, we see a positive and significant parameter estimate. We continue our discussion of this parameter estimate with respect to hypothesis 2P, below.

Hypothesis 3EC speaks to the notion that clustering is associated with "echo chambers" in social media. It is intended to represent the notion that people in clustered positions (those whose followers and followees also follow each other), may be even more likely than those in unclustered positions to tweet at a similar political slant to their network neighbors. Table 4, model IV reports a regression coefficient of 0.030 for the interaction between clustering and incoming slant. Therefore we reject the null for Hypothesis 3EC, as we do find evidence that people in positions of high clustering tweet more similarly to the people they follow than people in positions of low clustering. However, we must note that the effect of clustering, while statistically significant, is only of modest magnitude. A one-standard deviation increase in clustering coefficient would only result in a predicted increase in outgoing slant from 0.724 times the incoming slant to 0.754 times the incoming slant.

Finally, it should be noted that the models in Table 4 all have low $R^2$. Incoming slant predicts outgoing slant, but only with a lot of error. This large error is itself a sort of evidence against the idea of widespread echo chambers. If the mean difference between outgoing slant and incoming slant is substantial, as we see for the whole population, then clearly people are not only being exposed to views they would have tweeted themselves.

### 4.1.2 Polarization

The theory of polarization says that the average incoming slant should be more centrist than average outgoing slant, because people receive some information (that they disagree with) from the other side of the political spectrum. Thus, Hypothesis 1P says that the estimated parameter for incoming slant's effect on outgoing slant should be greater than one. Since the parameter in our models is at most approximately 0.72 (0.75 for 2017 data) we fail to reject the null for Hypothesis 1P for the whole population.

Although the main effect of incoming slant on outgoing slant is less than one, the parameter estimate for the interaction between the standard deviation of incoming slant and mean of incoming slant is positive and significant: the greater the exposure to diverse views, the more extreme outgoing slant is relative to incoming slant. Thus we reject the null for Hypothesis 2P. Although the population as a whole does not show evidence of polarization, this result suggests that there is some heterogeneity, and those that are exposed to more diverse viewpoints could be seen as more polarized by this measure. People exposed to a narrow range of slant tend on average to post more moderately than what they read, but the wider the range of slant that a user is exposed to, the less moderate they are. Our analysis of the network core, below, returns to this issue.

### 4.1.3 Mainstreaming

Hypothesis 1M considers the relationship between incoming and outgoing political slant from the perspective of mainstreaming theory. Although a strong version of the theory would say that the average Twitter user tweets at the same level of slant as the mean of the whole population, our Hypothesis 1M recognizes that people do not necessarily know where true political center is and says only that people tend to tweet more centrist material than the material they read in their own newsfeeds. As already stated, the estimated coefficient from Table 4 ranged from 0.67 to 0.72 (or 0.59 and .75 on models fit on 2017 data – see Appendix A), statistically significantly less than 1, and greater than 0. In other words, we reject the null for hypothesis 1M. Overall, Twitter accounts do tend to tweet more centrist material than the material posted by the accounts they follow, but not necessarily at or near the political mean of the population. As we explore below, however, there is heterogeneity in this population, and the average does not tell the whole story.

## 4.2 Beyond average behavior: macroscopic and subnetwork analyses

### 4.2.1 Core-periphery structure

Hypotheses 4H and 4P concern a core of highly followed users who are active in posting links to news stories. Hypothesis 4H predicts a larger estimated parameter for the effect of incoming slant on outgoing slant within the core (as compared to the periphery), and hypothesis 4P states that the higher the standards used to define the core, the higher that parameter will be. The parameter for the interaction between outdegree and incoming slant in Table 4 already provides suggestive evidence in favor of these hypotheses: the more followers a user has, the greater the expected slope of outgoing slant with respect to incoming slant.

Tables 5, 6, and 7 summarize more direct evidence relevant to these hypotheses: the estimated parameter for the effect of incoming slant on outgoing slant is reported for different definitions of the core with respect to both outdegree and news posting activity. When we consider only those tweets from inside the core, the maximum parameter estimate that we find is 1.086; when we consider all tweets from all sources, we find an even higher parameter: 1.172.

All specifications we tested for the core yielded a higher parameter for the effect of incoming slant on outgoing slant than the one we found in our study of the whole population in Table 4. We confirmed the statistical significance of these comparisons in unreported dummy variable regressions ($p$ was within machine precision of zero for all models); we thus reject the null for Hypothesis 4H.

As for Hypothesis 4P, there is a clear pattern evident in Tables 5 through 7: the more restrictive the definition of the core, the higher the estimated parameter for the effect of incoming slant on outgoing slant. In both tables, the higher the quantile of degree used as a threshold for core membership, the greater the estimated parameter. The magnitude of the effect of raising the quantile threshold of news posting activity is smaller than that for degree and in Table 5 is generally highest at the 90$^{th}$ quantile of news posting in each column, except the last column, corresponding to the strictest definition of the core. In this right-most column of both tables, the maximum parameter estimate is found when we define the core as consisting only of those individuals who are above the 95$^{th}$ percentile for both outdegree and number of news items posted. We therefore reject the null for hypothesis 4P: the stricter the definition of what constitutes the news-centric core, the greater is the effect of incoming slant on outgoing slant.

Table 5: Estimated parameter for incoming slant in regression of outgoing slant on incoming slant for news-centric core, with different definitions of which nodes belong to the core, considering only communication within the core

| | | Quantile of outdegree | | | | |
|---|---|---|---|---|---|---|
| | | 75th | 80th | 85th | 90th | 95th |
| Quantile of news posting | 95th | *0.9522 | 0.9731 | 0.9929 | 1.0099 | *1.0863 |
| | 90th | *0.9630 | 0.9804 | 1.0029 | 1.0281 | *1.0802 |
| | 85th | *0.9500 | *0.9689 | 0.9867 | 1.0196 | *1.0623 |
| | 80th | *0.9451 | *0.9622 | 0.9789 | 1.0162 | *1.0659 |
| | 75th | *0.9236 | *0.9433 | *0.9663 | 0.9988 | *1.0447 |

Note: * indicates 95% confidence interval for the mean does not contain 1.0

Table 6: Estimated parameter for incoming slant in regression of outgoing slant on incoming slant for news-centric core, with different definitions of which nodes belong to the core, considering all tweets.

| | | Quantile of outdegree | | | | |
|---|---|---|---|---|---|---|
| | | 75th | 80th | 85th | 90th | 95th |
| Quantile of news posting | 95th | *1.0362 | *1.0594 | *1.0878 | *1.1078 | *1.1723 |
| | 90th | *1.0294 | *1.0452 | *1.0714 | *1.1016 | *1.1477 |
| | 85th | 1.0072 | *1.0284 | *1.0505 | *1.0874 | *1.1270 |
| | 80th | 0.9944 | 1.0142 | *1.0370 | *1.0811 | *1.1315 |
| | 75th | *0.9613 | 0.9848 | 1.0093 | *1.0573 | *1.1093 |

Note: * indicates 95% confidence interval for the mean does not contain 1.0

Table 7: Number of accounts in news-centric core, with different definitions of which nodes belong to the core.

| | | Quantile of degree | | | | |
|---|---|---|---|---|---|---|
| | | 75th | 80th | 85th | 90th | 95th |
| Quantile of news posting | 95th | 6332 | 5489 | 4553 | 3478 | 1956 |
| | 90th | 11003 | 9427 | 7708 | 5705 | 3157 |
| | 85th | 15929 | 13480 | 10799 | 7866 | 4303 |
| | 80th | 20882 | 17473 | 13805 | 9879 | 5347 |
| | 75th | 30435 | 24973 | 19392 | 13513 | 7034 |

### 4.2.2 Polarization in the core

Tables 5 through 7 show that the more restrictive the definition of the network core, the higher the parameter estimate for the relationship between incoming slant and outgoing slant. For moderately restrictive definitions of the network core (for example, those accounts with greater than the 85$^{th}$ percentile of outdegree and 90$^{th}$ percentile of news items posted in Table 5) the parameter estimates for incoming slant are not significantly different from 1.0. We therefore would be able to reject the null for Hypothesis 1EC – that the mean political slant of news sources in tweets by individuals is statistically indistinguishable from the mean political slant of the tweets that they receive from the people they follow – for the news-centric core thus moderately defined. However, for the most restrictive definitions of the core the average outgoing slant is in fact more extreme than the average incoming slant, indicating not so much echo chambers, in which we would expect people to be reading and tweeting at the same political slant, but rather a tendency to polarization, in which we see people reading more centrist material on average than what they tweet themselves.

What emerges is a more nuanced picture of the whole. The vast majority of Twitter accounts that post news items do not post many of them, have a moderate number of followers among other news-posting accounts, and tend to post news items from more centrist sources than what they read themselves. On the other hand, a small minority of Twitter accounts constituting the network core posts relatively many news items from more politically extreme news sources than those in their own news feeds.

This is not to say that the core only posts material from the political extremes or that the periphery only posts centrist material, simply that on average the core posts more extreme material and the periphery posts more centrist material *than the accounts they follow*. Our results also do not support the extrapolation that the centrist tendency of accounts in the periphery is due to a tendency of following more extreme accounts in the core. We regressed outgoing slant on incoming slant after excluding core accounts and the tweets originating from those accounts (in the manner of Table 5, but for the periphery rather than for the core). After thus removing the effects of the core from the periphery, the estimated parameter for incoming slant's effect on outgoing slant was 0.7030, only slightly higher than the estimate for the complete data from Table 4, model I.

### 4.2.2 Correspondence of community structure and political slant: core

Is the Twitter follower network organized according to the political slant of its nodes? The theories of echo chambers and polarization would predict yes, while the theory of mainstreaming would predict no. Here we make several comparisons between permutations of nodes based on slant to those deriving from the patterns of ties alone using community discovery algorithms. Figures 2 and 3 visualize the adjacency matrices of the core and typical users subgraphs according to the community discovery-based and slant-based permutations of the nodes. The following paragraphs quantify these comparisons.

Section 3.4.3, above, describes the matrix probability model of which we calculate the likelihood on the algorithmic and slant partitions. In short, the likelihood of this model is calculated by pointwise multiplication of $Z$ (see above) with some permuted adjacency matrix $P$ and will be high to the extent that a given permutation concentrates tie weight toward the diagonal of a matrix. Critical values for differences in likelihoods were determined computationally.

The likelihoods of the diagonal gradient model under the four permutations of the core subgraph are presented in Table 8. Critical values are also given, which represent the 95$^{th}$ percentile of the log likelihood expected when 5% of nodes are removed from their proper place in the permutation and placed in a worst-fit location in the permutation. Any log likelihoods below a permutation's critical value can be considered worse descriptions of network structure.

Strikingly, the outgoing slant permutation is a much poorer fit to the diagonal gradient model than any of the other three permutations, and indeed the likelihood of Z given the outgoing slant permutation is significantly less than the other three according to our critical values, leading us to reject the null for Hypothesis 5Ma. Additionally, for the core, the likelihood of the incoming slant permutation is not less than the critical values of the Clauset, Newman and Moore (2004) and Laplacian eigenvector-based permutations. We fail to reject the null for Hypothesis 5Mb and find incoming slant to be an equivalently good description of network structure as standard community discovery algorithms. With evidence in favor of 5Mb but not 5Ma, our results are partially in accord with prior literature for the core.

### 4.2.3 Correspondence of community structure and political slant: periphery

The likelihood of the diagonal gradient model under the four permutations of the "moderate users" subgraph is presented in Table 9. Both of the slant permutations are worse than the critical value of the log likelihood for both of the community discovery algorithms. We therefore reject the null for Hypotheses 5Mc and 5Md: community discovery algorithms produce better descriptions of network structure than either incoming or outgoing political slant. For users outside of the core, our results contradict earlier studies: network structure is not well-described by the political slant of the users.

Table 8: log likelihood of *Z* under various permutations of nodes of the core

| | Log likelihood of probability model *Z* | Critical value to be considered worse than this permutation |
|---|---|---|
| Laplacian Eigenvector | -43063.8 | -55396.7 |
| Incoming Slant | -45213.0 | -57810.8 |
| Clauset, Newman and Moore | -45961.4 | -58396.4 |
| Outgoing slant | -72539.6 | |

Table 9: likelihood of Z under various permutations of nodes of the moderate users subgraph

| | Log Likelihood of probability model *Z* | Critical value to be considered worse than this permutation |
|---|---|---|
| Clauset, Newman and Moore | -27611.6 | -50379.4 |
| Laplacian Eigenvector | -35840.4 | -55604.4 |
| Incoming Slant | -79691.0 | -87822.2 |
| Outgoing slant | -94787.2 | |

Table 10: Summary of hypothesis tests by theory

| Hypothesis number | Hypothesis text | Main analysis: null Hypothesis rejected? | 2017 Robustness check: Null Hypothesis rejected? |
|---|---|---|---|
| *Homophily (H)* | | | |
| 1H | The mean political slant of news sources in tweets by individuals is significantly correlated to the mean political slant of the tweets that they receive from their followees. | Yes | Yes |
| 4H | The mean political slant of news sources in tweets sent by individuals in the news-centric core of Twitter users is closer to the mean political slant of the tweets that they receive from their followees than for people outside of the news-centric core. | Yes | |
| *Echo Chambers (EC)* | | | |
| 1EC | The mean political slant of news sources in tweets sent by individuals is the same as the mean political slant of the tweets that they receive from the people they follow. | No | No |
| 2EC | The smaller the standard deviation of political slant an account receives, the closer mean outgoing slant will be to mean incoming slant. | No | |
| 3EC | The greater the clustering around an individual, the closer the political slant in their own tweets is to the political slant in the tweets they receive from the people they follow. | Yes | |
| *Polarization (P)* | | | |
| 1P | The mean political slant of news sources in tweets by individuals is more extreme than the mean political slant of the tweets that they receive from the people they follow. | No | No |
| 2P | The larger the standard deviation of political slant an account receives, the more extreme outgoing slant is relative to incoming slant. | Yes | |
| 4P | The stricter the definition of what constitutes the news-centric core, the more extreme outgoing slant is, relative to incoming slant. | Yes | |
| *Mainstreaming (M)* | | | |
| 1M | The mean political slant of news sources linked to in an individual's own tweets is more politically centrist than the mean political slant in the tweets they receive from the people they follow. | Yes | Yes |
| 5Ma | Outgoing political slant is not a good description of the community structure of the network core. | Yes | |
| 5Mb | Incoming political slant is not a good description of the community structure of the network core. | No | |
| 5Mc | Outgoing political slant is not a good description of the community structure of the network periphery. | Yes | |
| 5Md | Incoming political slant is not a good description of the community structure of the network periphery. | Yes | |

# 5. DISCUSSION

## 5.1 Summary of empirical findings

Overall, our results are only partially consistent with theories of echo chambers, polarization and mainstreaming. A summary of hypothesis tests is presented in Table 10. Although small echo chambers may exist, we do not see clear evidence for them in the aggregate, and incoming slant only predicts outgoing slant with a lot of error. We do find evidence of homophily (outgoing slant is correlated with incoming slant), but also an average tendency to moderation and many points of contact among different points on the political spectrum (see slant-permuted matrices in Figures 2 and 3). We do see a polarized and active core in which network structure closely corresponds to political slant, but we also see a much larger (albeit much less active) generally moderating majority for which network structure is more weakly related to slant. A diagrammatic summary of the overall communication structure is in Figure 4. The widespread concern over polarization may be due to the over-representation of tweets originating in the core, constituting a sort of network paradox (Feld, 1991). As for mainstreaming, we do not find an absolute, but rather a relative tendency to political centrism. We also note that accounts outside of the core are tweeting across the political spectrum, which undermines a literal theory of a spiral of silence (Noelle-Neumann, 1974).

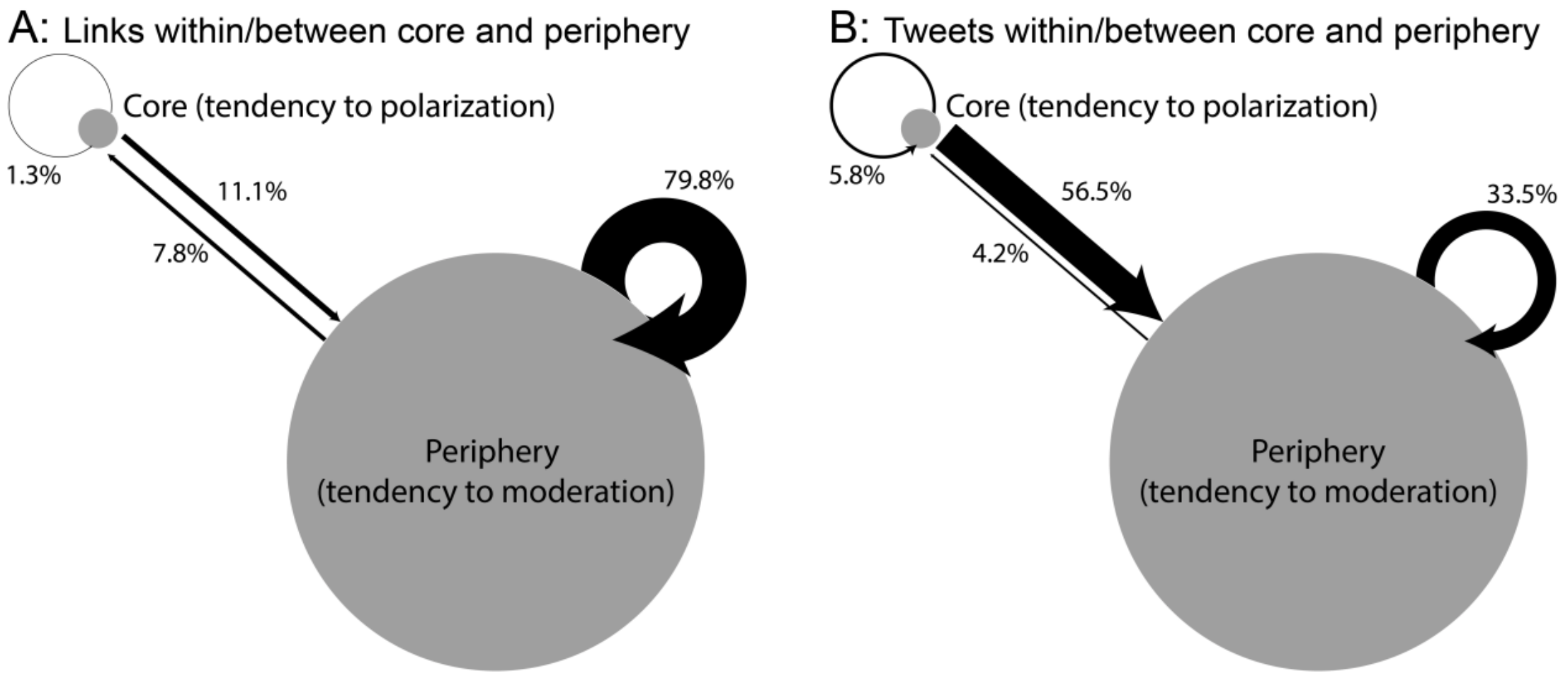

**Figure 4.**: Summary diagram of connectivity patterns, distinguishing core from periphery. Overall, there is a tendency to centrism, but a majority of tweets received originate in the network core, which has a tendency to polarization. Grey circles represent accounts in the network core and periphery. Circle size is proportionate to number of accounts. Arrows indicate percentage of total connectivity within and between core and periphery. **A:** arrow size is proportional to total number of follower-followee relationships in the full data set and labeled with a percentage (e.g. 79.8% of all links are within the periphery). **B:** arrow size is proportional to (an upper bound on) the number of tweets received in the full data set, calculated as number of tweets sent multiplied by the number of news-active followers (accounts that sent at least one link to a news site) that those tweets were sent to (e.g. only 33.5% of all tweets received were both sent and received by accounts in the periphery).

## 5.2 Broader implications

### 5.2.1 What is read versus what is said

In cross section, we find that communication patterns look very different when one looks at what is read (incoming information) instead of what is said (outgoing information). This may have substantial consequences for our understanding of influence in social networks, which typically only looks at expressed behavior (analogous to what is said in the context of this study). For example, in a network study of influence in the spread of a product, an individual's social media posts about that product could be interpreted as an expression of interest in the product or as the outward expression of desire to conform without any true interest in the product.

Additionally, we find that the relationship between what is read and what is said is strikingly different in the network core from outside of it: core accounts tend to position themselves in a more extreme position than what they are exposed to, while the typical account positions itself in a more moderate position. In the setting of influence in networks, it could well turn out that there is a similar regularity such that those within a core systematically express their preferences in an extreme or polarized manner, while those outside of the core systematically express their preferences in a moderate and dampened manner.

Because of this marked heterogeneity between core and periphery, it is necessary to study communicating systems as a whole as we seek to understand the technologically mediated crowd that is of increasing importance in our evolving economy and society.

### 5.2.2 The core versus the periphery in online communities: the "multiplex public"

Like other social networks, online communities have a core-periphery structure (Dahlander and Fredriksen, 2012; Collier and Kraut, 2012; Wasko, Teigland and Faraj, 2009) and are composed of individuals with shared goals and interests that communicate over the internet (Preece, 2000), in a self-organized manner consisting of voluntary participation and without formal organization (Dahlander and O'Mahoney, 2011). Our data could therefore be considered an online community of political discussion with liberal and conservative sub-communities, or alternatively, two overlapping communities in conflict with each other.

In general, prior research has treated membership in the core versus the periphery as essentially an issue of the *level of engagement* in the community. Some attention has been paid to how individuals end up in the core (Collier and Kraut, 2012; Dahlander and O'Mahoney, 2011; Johnson, Safadi and Faraj, 2015), and the sources of motivation for "heavy weight" participants in the core compared to "light weight" participants in the periphery of an online community (Haythornthwaite, 2009). Our results, however, reveal that those in the periphery are not only different from those in the core in terms of the amount of participation or reason for participation in the community, but indeed also in terms of the very nature of their information sharing behavior. Again, we find that on average, core members share links to more politically extreme news sources than the links they receive in their own timelines. Periphery members, on the other hand, are the opposite.

People tend to express themselves freely to the extent that the topic of conversation is consistent with their public or professional identity, and that their audience is homogenous (Marwick and boyd, 2010). For most people these conditions do not apply, since they use a personal (rather than professional or other narrowly constructed public identity) social media account to connect to multiple contexts and identities (Rainie and Wellman, 2012; Hampton, Lee and Her, 2011; Marwick and boyd, 2010). In other words, most people cannot assume that their followers also follow each other, which accords with the fact that the periphery of a social network is not highly interconnected within itself by definition (Borgatti and Everett, 2000). For people who both have a clear public identity and surround themselves with others with shared interests and goals – in other words, for members of the core – Marwick and boyd's (2010) conditions for free expression are met. This free expression could then be amplified by social influence (Centola and Macy, 2007; Shore, Bernstein and Lazer, 2015) and made more extreme by group polarization processes (Sunstein, 2002).

If individuals in the core and the periphery have different characteristic behaviors and social environments, then lumping them together under the single term "community" is insufficient. Instead, a new term is needed to describe this social structure that is most pervasive in our data. We offer the term "multiplex public" to describe the social structure that such typical users of social networking services inhabit. "Multiplex" refers to the multiple network layers (a work network, a school network, a friend network and so on) that come together to form the overall follower-followee network, and "public" emphasizes the environment that is neither a single cohesive community nor a disconnected crowd, but in which individuals are still visible to sparsely-connected others.

We suggest that this multiplex public has received less attention in the past in part because it has not been an obvious source of peer production. Because of their economic consequence, online communities and crowds have been obvious and important to researchers in and around the disciplines of management. Now, as data science uses digital traces for all manner of social scientific and business intelligence purposes, we should also acknowledge the significance of this prominent social structure and identify the ways it diverges from cohesive groups and network cores in future research.

**5.2.3 Research methods**

Network research nearly always faces a boundary definition problem (Laumann, Marsden and Prensky, 1989): the researcher must define who is in and who is out of the research data. As a matter of convenience, this often means selecting nodes on the basis of their activity; in the case of political slant, prior work has sampled people to study on the basis of their obvious political partisanship (Adamic and Glance, 2005; Conover, et al, 2011; Bakshy, Messing and Adamic, 2015; Barbera et al.,2015). While all of these studies go to some lengths to account for their data collection strategy, at a certain level they cannot fully escape the fundamental limitations that come with sampling on clearly political activity or affiliation. That partisans are polarized does not imply that social media users in general are polarized.

The implications for future research on social media are clear: the behavior of members of the core is not representative of people outside of the core. Networks constructed by choosing obviously relevant individuals (because they post a lot about the research topic, for example) are likely to consist only of the network core and leave out the more representative (in terms of ordinary users) periphery.

### 5.3 Limitations

Although our data is broadly representative in terms of its inclusion of typical Twitter users, our coverage consists of a cross-section in a non-election year. This means that we cannot speak to issues of influence or other dynamic processes on or of networks – only the cross-sectional organization of Twitter. More importantly, however, is the fact that our data was collected from a relatively "typical" period of time: 2009 was not an election year and September 10th-23rd (the data collection window) did not contain any major news stories[9] that might spark an increase in partisan conflict. If the data were collected at an atypically polarized time, we may have observed different results. Finally, Twitter was 3 years old when the data were collected, so while it no longer was only the home of early adopters, it had not yet gained the reach and user base that it has today. It is impossible to say for certain how this might affect results if this study could be exactly replicated with current data. To at least partially address these concerns we conducted robustness checks with current (2017) data that do agree with our main results and are presented in Appendix A.

A second set of limitations comes with our use of Gentzkow and Shapiro's slant scores. Although they cover over 95% of all direct news browsing and an even higher percentage of exposure to news on social media, we do not cover all sources of news. We cannot rule out the possibility that there are echo chambers built around the sharing of news from sites representing a tiny minority of news exposures, including those from hate sites. Indeed, if there were a total absence of such phenomena at the fringe, it would be surprising. However, this doesn't affect our results, which characterize the vast majority of news exposures on Twitter. Our robustness checks in Appendix A use a broader set of news sources than Gentzkow and Shapiro, and are intended to at least partially allay these concerns as well.

---

[9] See http://www.infoplease.com/year/2009.html#us

Our calculation of incoming slant is limited to those tweets that each account received from their followees in their timelines. We do not account for information received off of Twitter, or for any tweets than a user is exposed to through other means, such as searching for hashtags, reading "trending" topics, reading the tweets from users that one does not already follow (more recently, this list includes "promoted" tweets, that is, advertisements). We cannot say for certain how this might affect our results, but one possibility is that if people systematically searched for information from the opposite side of the political spectrum, it could result in undetected polarization, because their true incoming slant would be more centrist than we have measured it.

A relative strength of our study is the inclusion of nearly every account that shared a news link during the collection period and thus had a measurable behavior. But what about those accounts without any measurable behavior with respect to our slant measure? A rough estimate is that our population of ~208,000 accounts that tweeted and received a news link represents only about 2.5% of monthly active users[10]. It is hard to say how our results would relate to the 97.5% of accounts that did not tweet any news links whatsoever during our data collection window. Given the diversity of interests – especially popular culture – represented on Twitter, it is likely that for quite a number of these excluded individuals, their twitter behavior is simply orthogonal

[10] We are not aware of a reliable estimate for the number of monthly active users for the data collection period. Extrapolating backwards from information in Twitter's filing of form S-1 with the United States Securities and Exchange Commission on October 3, 2011, it would appear that there were approximately 8 million monthly active users in the United States at the time of our data collection

to the issues we are studying. However, other individuals may have participated in political communication through other means than sharing a news article; what can we say about these accounts? We found that the fewer news links an account sent out, the more centrist they tended to tweet relative to the information they received. Extrapolating from this, one might conclude that those who sent no news links would be even more mainstream (if that were measurable). On the other hand, there could well exist very political individuals who simply to not share links to news stories, and instead write their own opinions in their own words. If these users are somehow systematically different from our study population, then we may be missing some important part of the broader phenomenon of information diversity on Twitter.

Finally, we study incoming versus outgoing political slant for Twitter accounts. It is not warranted to extrapolate our results without major caveats to conclusions about the true ideology of individual people. There is also the possibility that some subset of users share articles they disagree with, which would complicate interpretation of our results. We conducted exploratory analyses of this issue, and could not find much evidence of this effect in our data set. A complete answer to this question, however, would require a separate full research project and is outside of the scope of this article. Accounts also do not necessarily correspond to individuals; rather, they can also be run by organizations or algorithms. Apart from some very obvious cases, it is hard to distinguish organizations from individuals on Twitter. Individuals whose professional work involves participating in public discussion of news (reporters, academics, authors, public figures and so on) can have accounts that promote those individuals' "personal brands" and in some cases may even be run by a private social media team. The line between individuals and organizations is blurry in the public forum of Twitter. If we consider, for convenience, the network core to have a higher density of organizational accounts, then it could be slightly reassuring that repeating our analyses with only accounts in the periphery produced almost identical estimated parameters for the effect of outgoing slant on incoming slant as those in Table 4. As for algorithms, robustness checks in Appendix A on a small set of accounts are suggestive that our main results hold after excluding bots.

### 5.4 Conclusion

By using data representative of the whole population of Twitter users, we were able to reconcile apparently contradictory theories of diversity of information sharing on Twitter. The aggregate picture cannot be described as just a collection of echo chambers on the one hand, or a clear pattern of mainstreaming on the other. Rather, with elements of both tendencies, we instead see a whole system comprising a vast moderating majority – a multiplex public – with a polarized two-part community at its core. Predicted behavior depends on which part of the system you are looking at, but on average, Twitter accounts post more centrist information than they receive in their own timelines, undercutting the prevailing narrative of the social media echo chamber (e.g. Schmidt, et al, 2017). Instead, the widespread perception of such polarization may be the result of a network paradox, in which the behavior of nodes with a high degree is mistaken to be typical (Feld, 1991).

**APPENDIX A: Robustness check on recent data**

**SUMMARY**

We conducted an extensive robustness check on recent (Feb-March, 2017) data. We calculated our own measure of slant by analyzing patterns of co-following among people who follow news sources directly. We then assembled a sample of Twitter users and analyzed incoming versus outgoing slant. To the extent that we were able to conduct analogous analyses, our main result stands with current data.

**A1: CALCULATION OF SLANT SCORE**

In our main analysis we had an externally validated measure of slant that corresponded to the same period in time that our data was from. We felt that the measures from 2009 were much too old to rely on for current data, and that many important news sources (for example, Wikileaks and fivethirtyeight.com) had arisen in the intervening years. We thus devised our own measure that was closely related in principle to the measure from Gentzkow and Shapiro for the 2009 data. In particular, we use their notion that bias is captured by audience segregation. In their methods, a news site was deemed conservative (i.e. having a positive slant) if it was relatively heavily-read by conservative people and had relatively few liberal people among its audience. News sites with centrist scores (near zero) were read by both conservative and liberal people. We adapt this idea to Twitter data as detailed here.

We began by identifying news sources to include in our slant calculations. We created a list of 186 different sources, including most of the sources from the 2009 data (we excluded sources without at least 10,000 followers on Twitter, as well as blogtalkradio.com, because it is a platform for others to share their content and at this point in time does not have a predictable slant). We then extended the original list by searching for curated lists of political news sources and added all of those as well (limiting our sources to those that had at least 10000 followers on Twitter). We also separated CNN politics from the rest of CNN, the Wall Street Journal opinion section from the rest of the Wall Street Journal, and the Politics and Black Voices sections from the rest of the Huffington Post, as those divisions are thought to have a different slant from the rest of the reporting on those sites, according to multiple comments we read.

We then identified the main twitter handle of each of these news sources and used the Twitter API to download the lists of direct followers of these news sources. This step took approximately 7 weeks of continuous querying of the API in order to collect 330,000,000 follower relationships to our list of 186 news sources (these 7 weeks were divided across multiple developer accounts so the actual time spent collecting data was closer to two weeks). Collecting data at this scale necessitated building a collection script that was robust to undocumented behaviors of the API.

Having collected the lists of followers, we then calculated a 186x186 affinity matrix, the entries of which represent how much each pair of news sources overlapped in terms of followers. For example, at the time of data collection, the New York Times had approximately 33,000,000 followers on Twitter, and the Blaze (Glenn Beck) had approximately 616,000 followers; about 165,700 people followed both sites, which is 26.9% of the size of the smaller audience. We thus use 0.269 as the entry in our affinity matrix for the strength of the connection between the New York Times and the Blaze. In general, denoting our affinity matrix as A,

$$A_{ij} = \frac{|\text{Followers}_i \cap \text{Followers}_j|}{\min(|\text{Followers}_i|, |\text{Followers}_j|)}$$

where |X| denotes the cardinality of set X.

Ultimately, our slant scores are calculated from this raw affinity matrix. However, analogous to Gentzkow and Shapiro's method, we needed to first identify subsets of news sources that were clearly embedded in a partisan cluster such that we had a basis for approximating partisan audiences. The intention behind this is to use co-following of sites in the conservative cluster as an indicator of conservatism, and co-following of sites in the liberal cluster as an indicator of liberalism.

However, as others have recently pointed out (e.g. Benkler et al.'s preliminary report on their research here: http://www.cjr.org/analysis/breitbart-media-trump-harvard-study.php), the conservative news cluster is clearly distinct, while the liberal news cluster is more integrated with the mainstream news cluster (that is, people who follow news sources with a liberal slant also follow mainstream news sources more than people who follow news sources with a conservative slant do), as Figure A1 illustrates. Because of this overlap, community detection algorithms run on the raw affinity matrix tend to divide the news site co-following network into only two, rather than three communities. However, in order to calculate an analogous measure of slant to the one in our main analysis, we needed to identify which sites were in the less distinct smaller cluster of clearly liberal-leaning sites.

As an intermediate step, we analyzed a scaled version of the matrix rather than the raw matrix to emphasize the community structure and provide a clearer signal of partisan separation. In particular, we reasoned that some co-following is to be expected if only based on random chance. For example, for a very small niche news site, it wouldn't be terribly surprising if a substantial fraction of its audience also followed CNN or the New York Times. Intuitively, the smaller the first site, the less surprising it would be if a large fraction of its audience also followed larger news sites as well. Thus, we scale the entries in the affinity matrix by the ratio of the natural logs of the number of followers for each news site. This penalizes co-following with sites that are much larger and emphasizes co-following that is more unexpected with respect to relative audience size. After thus scaling the matrix, we removed all edges with scaled weight less than 0.3 and trimmed nodes with resulting degree less than 5. These latter steps emphasize stronger connections and greater embeddedness in communities. We then ran the spin glass-based community detection algorithm (J. Reichardt and S. Bornholdt: Statistical Mechanics of Community Detection, Phys. Rev. E, 74, 016110 (2006)) on this pre-processed network to find three clear communities.

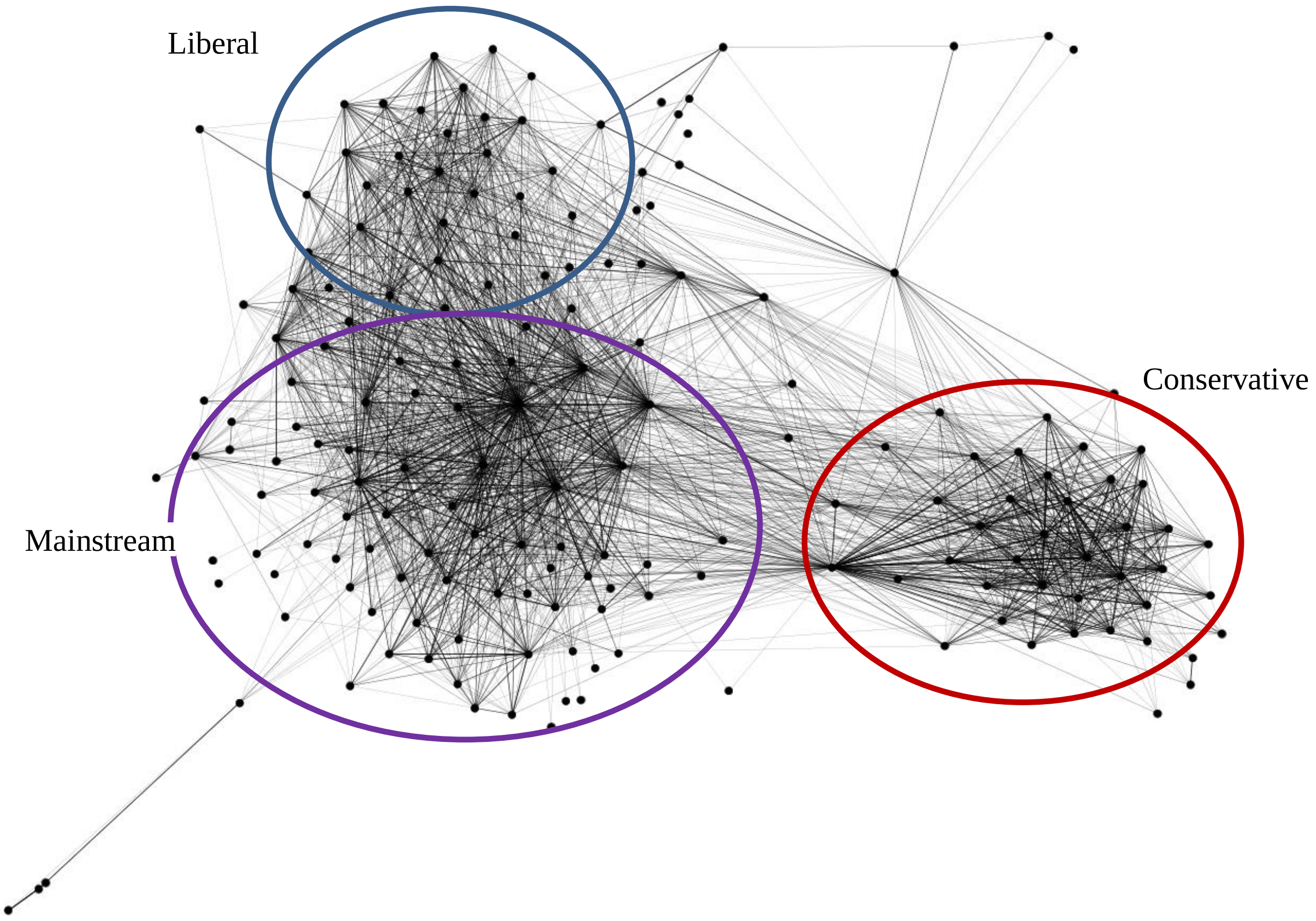

Figure A1: Visualization of raw affinity matrix (note that for the purposes of this visualization only, weak ties (those less than 0.25) are omitted).

Having thus identified nodes belonging to a liberal and a conservative cluster, we return to the raw affinity matrix to calculate slant scores (that is, the pre-processed version of the matrix was only used to detect membership in the three clusters). For each news site, we calculated the mean co-following between that site and sites in the liberal cluster divided by the mean co-following with all sites in the network and multiplied that by negative one to represent the liberal-leaning tendency. We then added that to the mean co-following between that site and sites in the conservative cluster divided by the mean co-following with all sites in the network to represent the conservative leaning slant. This sum represents the estimated slant for each site about which we collected data.

$$Slant_i = -1 * \left( \frac{\frac{1}{|liberal|} \sum_{\substack{1 \le i \le n, \\ j \in liberal}} A_{ij}}{\frac{1}{n} \sum_{i,j,j \ne i} A_{ij}} \right) + 1 * \left( \frac{\frac{1}{|conservative|} \sum_{\substack{1 \le i \le n, \\ j \in conservative}} A_{ij}}{\frac{1}{n} \sum_{i,j,j \ne i} A_{ij}} \right)$$

Our resulting estimated slant scores correspond fairly well to other published measures of slant, such as Pew Research Center's 2014 report on political polarization (http://www.journalism.org/2014/10/21/political-polarization-media-habits/pj_14-10-21_mediapolarization-08/). The correlation between our slant scores and Gentzkow and Shapiro's 2009 scores is 0.79; the two news sources with difference between our score and Gentzkow and Shapiro's scores are the hotairblog.com and breitbart.com. Both were relatively new at the time of Gentzkow and Shapiro's data collection, and Breitbart has been documented to have evolved substantially since 2009[11].

[11] see, for example, "From agitator to enforcer: The evolution of Breitbart" at https://www.washingtonpost.com/video/politics/from-agitator-to-enforcer-the-evolution-of-breitbart/2017/02/19/9714b4ce-f3b8-11e6-9fb1-2d8f3fc9c0ed_video.html

Because liberals are more likely to be exposed to mainstream news sources, a truly unbiased source would not necessarily have a slant score of zero. Note, for example, that C-SPAN, a source that largely presents footage of congress without commentary, has an estimated slant score of approximately -0.29 on the scale for our 2017 data (which is on a different scale than Gentzkow and Shapiro's slant scores). Estimated slant scores are plotted in Figure A2 and reported in full in Table A2, below.

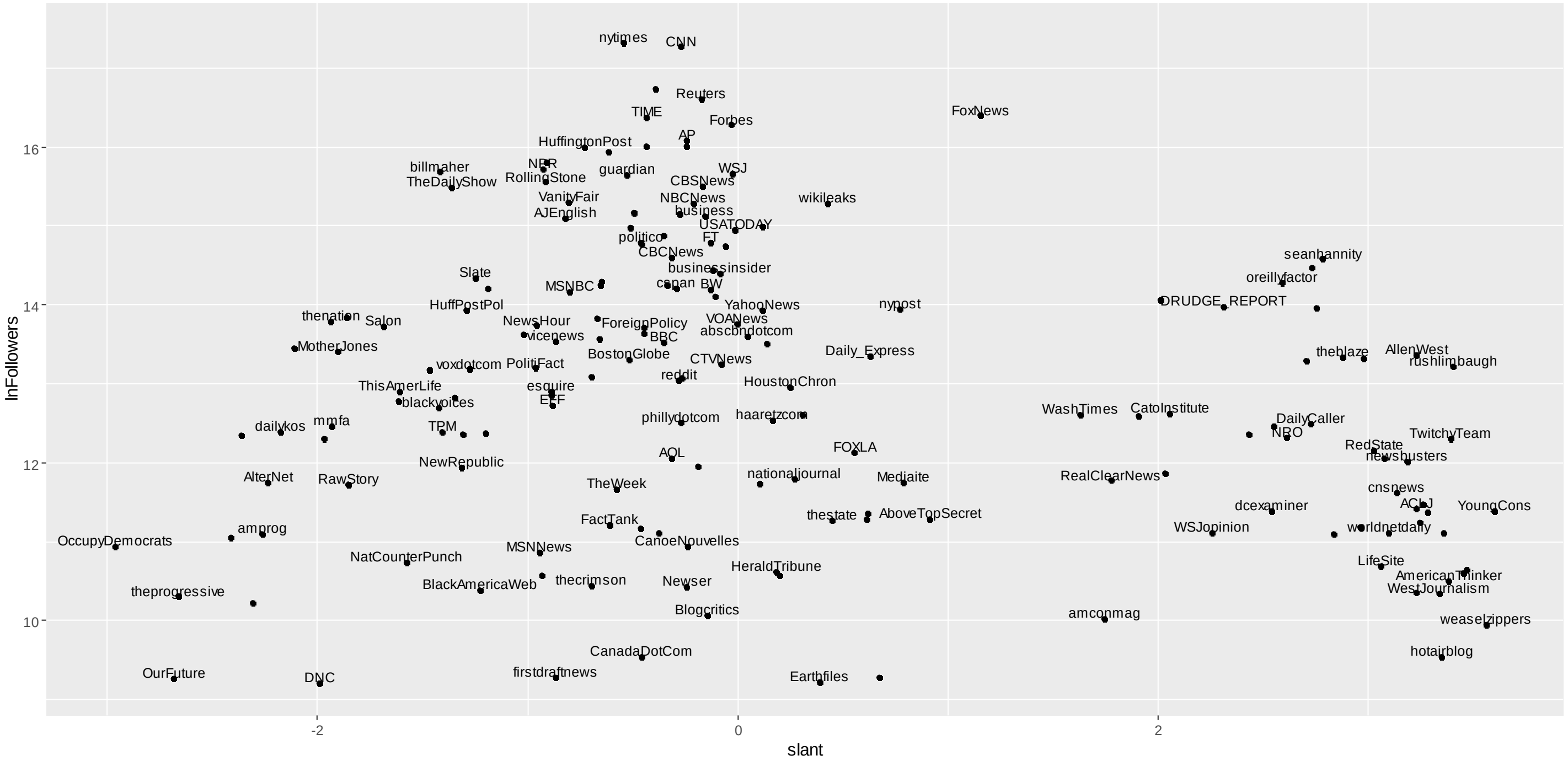

Figure A2: estimated slant scores for robustness check. The logarithm of the number of followers is on the vertical axis.

**A2: COLLECTION OF DATA ON INDIVIDUAL ACCOUNTS**

As indicated above, only a minority of Twitter accounts actually tweet links to news articles; sampling accounts at random (to the extent that is possible) to find those that both received and sent tweets containing links to news articles would be prohibitively expensive in terms of time. In order to find accounts more efficiently, we instead began by searching for tweets containing links to news stories, find the accounts that sent these tweets, and then searching among the followers of these accounts until we found those that also tweeted at least one link. Using this method, we obtain the full recent timeline (up to 3200 tweets, including only those less than or equal to two weeks old) for both the upstream followee and the downstream follower. We use the mean slant of the followee's tweets as a predictor variable and the mean slant of the follower as the dependent variable in regressions that we use as a robustness check of our main result.

Because of the extremely slow process of finding accounts that fit our criteria, our largest set of data contains only the slant of this single followee, rather than the entire incoming timeline of a given follower to predict that follower's outgoing slant. This obviously introduces substantial error (including attenuation bias) that was not present in our initial analysis, in which we had the entire incoming timeline for every account in our dataset. We have two reasons to believe that the error introduced by this method may not be as problematic as it sounds. First, the followees that we find by sampling on tweets (i.e. searching for tweets with the search API) are likely to be more representative of the timeline than a followee picked at random; this is because sampling on tweets is more likely to find especially active tweeters, whose tweets make up a majority of follower's incoming timelines. Second, for a subset of data, we also collected expanded incoming timelines and found generally consistent regression results.

Our full algorithm for collecting individual-level data is below, but before detailing that, we begin by defining a subroutine – "PROCESS URLS" -- used repeatedly throughout our data collection. For example, we frequently needed to determine the slant of the most recent 3200 tweets sent by a given account. URLs in tweets are always shortened links – that is, the original full URL is converted into a shorter version by Twitter (using the t.co domain) and possibly other shortening services such as bit.ly. This means that the links in any tweet data we collected are totally opaque with respect to where they point.

PROCESS URLS: Given a set of tweets containing URLS, we begin by discarding any that are more than two weeks old. We then follow each link until there are no further redirects and collect the URL at the final destination. We then strip away protocol and server names ("http://" or "https://" and "www" or other leading information) and match to the list of news domains we created above, discarding any tweets without links to a news source on our list.

Collection Algorithm

We query the search API for Tweets containing hyperlinks. We then PROCESS URLS (as above) for these tweets and call remaining tweets "seed tweets". For each seed tweet, we look up the account that sent the tweet, designating this account as "followee." If the followee has greater than 10,000 followers, stop and move on to the next tweet without recording data Query the REST API for the most recent 3200 tweets sent by the followee and then PROCESS URLS and calculate the average slant of news tweets sent by the followee. We then query the API for the list of accounts that follow the followee. Until we find a follower that sent a news link, loop over the following steps: {randomly select one follower, query the API for the most recent 3200 tweets sent by that follower, PROCESS URLS} (if no followers tweeted a news link, stop and move on to the next seed tweet). Calculate the average slant of the follower. Store data and move on to the next seed tweet.

**Further robustness checks**

The above collection algorithm allows us to regress a follower's outgoing slant on the outgoing slant of an account that they follow. Two substantial concerns with this approach are as follows: (1) in 2017, the number of "bots" (accounts that are not run by a human, but rather by a piece of computer code) has reportedly increased on Twitter and (2) it is unclear how representative one followee account would be of the whole incoming timeline (notwithstanding our note that because the followee is sampled on tweet activity, they are likely to be more representative than the average followee). To check the magnitude of any effects caused by these issues, we conducted further robustness checks on a subset of data.

**Bot check**: the Truthy team at Indiana University recently released an API to their "Bot or not" service (http://truthy.indiana.edu/botornot/). Essentially, this is a machine-learning based classifier of twitter accounts that produces a score, ranging from zero to one, with one being highest certainty that the account in question is a bot. We took a subset of followers from our main collection algorithm and used the botornot API to provide a bot score. We then kept only those accounts that scored less than 0.4.

**Expanded timeline**: with the subset of accounts unlikely to contain many bots, we then expanded the number of followees in the "incoming slant" calculation. We did not attempt to include all followees, because of feasibility constraints imposed by Twitter's rate limits and the requirement to PROCESS URLS. Instead, using the insight from the main analysis of 2009 data that heavier tweeters tend to tweet less centrist material we obtained timelines from only the top 20 heaviest tweeters from among each follower's followees. This also has the advantage that the more an account tweets, the more representative of the whole incoming slant they are, all else equal.

**A3: RESULTS**

Results from these robustness checks are presented in Table A1. Our main result, that on average outgoing slant is more moderate than incoming slant stands with these current data. These new data are much less complete than our main analysis and necessarily introduce some sampling bias of unknown magnitude. However, they were collected at a time of greater maturity of the Twitter platform and a time of much greater apparent political polarization – during the first months of the Trump administration – than the data for the main analysis.

Because of the different approaches to data collection, a detailed interpretation of differences between parameter estimates from 2009 to 2017 is generally not warranted. However, one substantial difference is attributable to the sampling approach and we note it here. The parameter for the interaction between incoming slant and number of followers was large and significant in our main analysis, but insignificant here. This is almost certainly because we only collected data for accounts with fewer than 10,000 followers for this robustness check; thus there is much less variation in this variable by construction.

Table A1 Relationship between incoming and outgoing political slant

| | DV: Mean slant of sites in outgoing tweets | |
|---|---|---|
| | I | II[‡] |
| Mean slant, sites in incoming tweets[†] | 0.586*** | 0.751*** |
| | 0.010 | 0.032 |
| *ln*(Count of outgoing tweets) | 0.064 *** | 0.007 |
| | 0.010 | 0.034 |
| *ln*(# followers) | 0.048 *** | 0.094 |
| | 0.014 | 0.048 |
| *ln*(# followers) ÷ *ln*(#followees) | -0.027 | -0.016 |
| | 0.014 | 0.048 |
| Incoming slant x | 0.154 *** | 0.093** |
| ln(Count outgoing tweets) | 0.010 | 0.033 |
| Incoming slant x | 0.009 | 0.038 |
| *ln*(# followers) | 0.014 | 0.048 |
| Inc. slant x *ln*(#followers) ÷ *ln*(#followees) | 0.036 * | 0.011 |
| | 0.017 | 0.057 |
| Intercept | -0.049*** | -0.009 |
| | 0.010 | 0.033 |
| # of Twitter accounts | 5966 | 445 |
| Adjusted $R^2$ | 0.419 | 0.589 |

Notes: standard errors are printed below parameter estimates.
***: p< 0.001; **: p< 0.01; *: p < 0.05
Except for mean incoming slant and mean outgoing slant, all variables are centered and scaled to unit variance
[†] For model I, incoming tweets originate from one sampled followee. For Model II, incoming tweets originate from the top 20 most-frequent tweeters among that account's followees
[‡]Model II includes a subset of accounts that have a low probability of being bots and have fuller incoming slant information

| Table A2: estimated slant scores for 2017 data | | |
|---|---|---|
| Twitter handle | URL | estimated slant |
| ABC | http://ABCNews.com | -0.241 |
| AboveTopSecret | http://www.abovetopsecret.com | 0.916 |
| abscbndotcom | http://www.ABS-CBN.com | 0.048 |
| ACLJ | http://www.ACLJ.org | 3.232 |
| ACLU | http://www.aclu.org | -1.861 |
| AJEnglish | http://aljazeera.com | -0.823 |
| AllenWest | http://www.allenbwest.com | 3.229 |
| AlterNet | http://www.AlterNet.org | -2.238 |
| amprog | http://www.americanprogress.org | -2.267 |
| AmericanThinker | http://www.americanthinker.com | 3.384 |
| AOL | http://www.aol.com | -0.317 |
| AP | http://www.ap.org | -0.245 |
| azcentral | http://www.azcentral.com | 0.306 |
| BBC | http://www.bbc.co.uk | -0.355 |
| billmaher | http://www.billmaher.com/ | -1.415 |
| oreillyfactor | http://www.billoreilly.com | 2.591 |
| BlackAmericaWeb | http://www.blackamericaweb.com | -1.227 |
| Blklivesmatter | http://www.blacklivesmatter.com | -1.971 |
| Blogcritics | http://blogcritics.org/ | -0.144 |
| business | http://www.bloomberg.com | -0.158 |
| BoingBoing | http://www.boingboing.net | -1.310 |
| BostonGlobe | http://bostonglobe.com | -0.517 |
| bostonherald | http://www.bostonherald.com | 0.617 |
| BreitbartNews | http://breitbart.com | 2.978 |
| businessinsider | http://businessinsider.com/ | -0.083 |
| BW | http://www.businessweek.com | -0.130 |
| BuzzFeedNews | http://www.buzzfeed.com/news | -0.657 |
| cspan | http://www.c-span.org | -0.294 |
| CanadaDotCom | http://www.canada.com | -0.456 |
| CatoInstitute | http://www.cato.org/ | 2.053 |
| CBCNews | http://www.cbc.ca/news/ | -0.313 |
| CBSNews | http://CBSNews.com | -0.169 |
| chicagotribune | http://chicagotribune.com | -0.446 |
| chicksonright | http://chicksontheright.com | 3.452 |
| CTmagazine | http://christianitytoday.com | 1.910 |
| HoustonChron | http://www.chron.com | 0.250 |
| CNBC | http://cnbc.com | -0.057 |
| CNET | http://www.cnet.com | -0.108 |
| CNN | http://www.cnn.com | -0.273 |
| CNNPolitics | http://cnn.com/politics | -0.648 |
| cnsnews | http://facebook.com/cnsnewscom | 3.137 |
| NatCounterPunch | http://counterpunch.org | -1.575 |
| crooksandliars | http://www.crooksandliars.com/ | -2.409 |
| csmonitor | http://www.CSMonitor.com | -0.372 |
| CTVNews | http://www.ctvnews.ca | -0.079 |
| DailyCaller | http://www.dailycaller.com | 2.728 |
| dailykos | http://www.dailykos.com | -2.176 |
| demunderground | http://www.democraticunderground.com | -2.227 |
| DNC | https://democrats.org/ | -1.990 |
| DRUDGE_REPORT | http://www.DRUDGEREPORT.com | 2.312 |
| Earthfiles | http://www.earthfiles.com | 0.390 |
| TheEconomist | http://www.economist.com | -0.390 |
| EFF | https://www.eff.org | -0.884 |
| esquire | http://www.esquire.com | -0.891 |
| MichaelSalla | http://exopolitics.org/ | 0.179 |
| Daily_Express | http://www.express.co.uk | 0.630 |
| YahooFinance | http://finance.yahoo.com | 0.138 |
| firstdraftnews | http://firstdraftnews.com | -0.865 |
| FiveThirtyEight | http://www.fivethirtyeight.com | -1.017 |
| Forbes | http://forbes.com | -0.028 |
| ForeignPolicy | http://www.foreignpolicy.com | -0.444 |
| FOXLA | http://www.foxla.com | 0.556 |
| FoxNews | http://www.foxnews.com | 1.157 |
| CanoeNouvelles | http://fr.canoe.ca/ | -0.239 |
| FreeBeacon | http://FreeBeacon.com | 2.960 |
| Snowden | https://freedom.press | -0.355 |
| FreeRepublicTXT | http://www.freerepublic.com | 0.176 |
| fpmag | http://frontpagemag.com | 3.287 |
| FT | http://www.ft.com/ | -0.127 |
| GallupNews | http://www.gallup.com | 0.613 |
| glennbeck | http://www.glennbeck.com | 2.754 |
| globeandmail | http://www.globeandmail.com | -0.336 |
| GOP | http://gop.com | 2.011 |
| haaretzcom | http://www.haaretz.com | 0.168 |
| seanhannity | http://hannity.com | 2.781 |
| HarvardBiz | http://hbr.org | -0.494 |
| HeraldTribune | http://www.heraldtribune.com | 0.182 |
| Heritage | http://heritage.org | 2.705 |
| hotairblog | http://hotair.com | 3.352 |
| HuffingtonPost | http://www.huffingtonpost.com | -0.730 |
| blackvoices | http://www.huffingtonpost.com/black-voices/ | -1.425 |
| HuffPostPol | http://www.huffingtonpost.com/politics | -1.291 |
| HumanEvents | http://www.HumanEvents.com | 3.228 |
| TheIJR | http://ijr.com | 2.036 |
| infowars | http://www.infowars.com | 2.430 |
| zittrain | http://www.jz.org | -0.930 |
| latimes | http://latimes.com/ | -0.454 |
| lessig | http://lessig.org | -1.347 |
| LifeNewsHQ | http://www.LifeNews.com | 3.075 |
| LifeSite | http://www.lifesitenews.com | 3.060 |
| MarketWatch | http://www.marketwatch.com/ | 0.119 |
| mashable | http://mashable.com | -0.614 |
| Mediaite | http://www.Mediaite.com | 0.788 |
| mmfa | http://mediamatters.org/ | -1.929 |
| PolitiFact | http://membership.politifact.com | -0.963 |
| michellemalkin | http://www.michellemalkin.com | 2.732 |
| MotherJones | http://www.motherjones.com | -1.906 |
| MoveOn | http://MoveOn.org/ | -2.365 |
| MSNBC | http://msnbc.com | -0.803 |
| nationaljournal | http://www.nationaljournal.com/ | 0.271 |
| NRO | http://www.NationalReview.com | 2.615 |
| NatureNews | http://www.nature.com/news | -0.654 |
| NBCNews | http://NBCNews.com | -0.210 |
| NewRepublic | http://www.newrepublic.com | -1.316 |
| MSNNews | http://news.msn.com | -0.944 |
| YahooNews | http://news.yahoo.com | 0.118 |
| newsbusters | http://www.newsbusters.org/ | 3.185 |
| Newser | http://www.newser.com | -0.243 |
| newsmax | http://www.newsmax.com | 2.834 |
| newsobserver | http://www.newsobserver.com | -0.188 |
| Newsweek | http://www.newsweek.com | -0.512 |
| NewYorker | http://www.newyorker.com | -0.908 |
| NPR | http://www.npr.org | -0.924 |
| nypost | http://www.nypost.com | 0.773 |
| nytimes | https://www.nytimes.com/ | -0.543 |
| OccupyDemocrats | http://www.OccupyDemocrats.com | -2.962 |
| OurFuture | http://www.ourfuture.org | -2.687 |
| NewsHour | http://pbs.org/newshour | -0.959 |
| FactTank | http://www.pewresearch.org/fact-tank | -0.612 |
| phillydotcom | http://www.philly.com | -0.270 |
| instapundit | https://pjmedia.com/instapundit/ | 3.248 |
| politico | http://www.politico.com | -0.460 |
| powerlineUS | http://www.powerlineblog.com | 3.429 |
| theprogressive | http://www.progressive.org | -2.662 |
| ProPublica | http://www.propublica.org/ | -1.467 |
| theprospect | http://prospect.org/ | -2.306 |
| PsychToday | http://www.psychologytoday.com | -0.697 |
| RawStory | http://www.rawstory.com | -1.856 |
| RealClearNews | http://www.realclearpolitics.com | 1.775 |
| reddit | http://reddit.com | -0.281 |
| RedState | http://www.redstate.com | 3.027 |
| Reuters | http://www.reuters.com | -0.175 |
| RollingStone | http://www.rollingstone.com | -0.915 |
| RSPolitics | http://www.rollingstone.com/politics | -2.026 |
| rushlimbaugh | http://www.rushlimbaugh.com | 3.406 |
| Salon | http://www.Salon.com | -1.686 |
| SELFmagazine | http://www.self.com | -0.268 |
| SFGate | http://sfgate.com | -0.889 |
| shadowproofcom | http://shadowproof.com/ | -1.744 |
| Slate | http://www.slate.com/ | -1.249 |
| TPM | http://www.talkingpointsmemo.com/ | -1.406 |
| technorati | http://technorati.com | -0.461 |
| amconmag | http://www.theamericanconservative.com | 1.745 |
| TheAtlantic | http://www.theatlantic.com | -1.190 |
| theblaze | http://www.TheBlaze.com | 2.883 |
| thecrimson | http://thecrimson.com | -0.700 |
| thedailybeast | http://thedailybeast.com | -0.668 |
| TheDailyShow | http://thedailyshow.com | -1.363 |
| gatewaypundit | http://www.thegatewaypundit.com/ | 3.360 |
| guardian | https://www.theguardian.com | -0.528 |
| thehill | http://www.thehill.com | -0.119 |
| MattWalshBlog | http://themattwalshblog.com/ | 3.284 |
| thenation | http://thenation.com | -1.938 |
| theolympian | http://www.theolympian.com | -0.554 |
| TPCblog | http://thepoliticalcarnival.net | -3.346 |
| trscoop | http://therightscoop.com | 3.469 |
| thestate | http://www.thestate.com | 0.452 |
| TheWeek | http://www.TheWeek.com | -0.576 |
| thinkprogress | http://www.thinkprogress.org | -2.112 |
| ThisAmerLife | http://www.thisamericanlife.org | -1.610 |
| TIME | http://www.time.com | -0.434 |
| TODAYshow | http://TODAY.com | -0.274 |
| Topix | http://topix.com | 0.573 |
| townhallcom | https://townhall.com/ | 3.262 |
| TreeHugger | http://www.treehugger.com | -1.616 |
| TwitchyTeam | http://www.twitchy.com | 3.395 |
| UPI | http://www.upi.com | 0.201 |
| USATODAY | http://www.usatoday.com | -0.016 |
| usnews | http://www.usnews.com | 0.107 |
| VanityFair | http://www.vanityfair.com | -0.808 |
| vicenews | http://www.vicenews.com | -0.866 |
| villagevoice | http://www.villagevoice.com | -1.202 |
| VOANews | http://voanews.com/ | -0.001 |
| voxdotcom | http://vox.com | -1.276 |
| dcexaminer | http://www.washingtonexaminer.com | 2.543 |
| washingtonpost | http://washingtonpost.com | -0.435 |
| WashTimes | http://www.washingtontimes.com | 1.632 |
| weaselzippers | http://www.weaselzippers.us | 3.564 |
| weeklystandard | http://www.weeklystandard.com | 2.555 |
| WestJournalism | http://www.westernjournalism.com | 3.341 |
| wikileaks | http://wikileaks.org | 0.428 |
| worldnewsdotcom | http://wn.com/ | 0.671 |
| worldnetdaily | http://www.wnd.com | 3.099 |
| WSJ | http://wsj.com | -0.023 |
| WSJopinion | http://wsj.com/opinion | 2.256 |
| YoungCons | http://youngcons.com/ | 3.602 |